\documentclass[11pt]{article}
\usepackage{jheppub}
\usepackage{graphicx}
\usepackage{booktabs}
\usepackage{multirow}
\usepackage{makecell}
\usepackage{amsmath}
\usepackage{amssymb}
\usepackage{xcolor}
\usepackage{tikz}
\usetikzlibrary{decorations.pathmorphing, decorations.pathreplacing, calc, shapes.geometric, arrows.meta, positioning, intersections}
\usepackage{pgfplots}
\pgfplotsset{compat=1.17}
\usepackage{url}
\newcommand{\MHVb}{\overline{\text{MHV}}}

\title{Unique loop integrands of planar $\mathcal{N}=4$ super Yang--Mills theory from soft and collinear factorization}

\author[a]{Dimitri Corradini}
\author[a]{and Johannes M. Henn}
\affiliation[a]{Max Planck Institute for Physics, Boltzmannstra\ss e 8, 85748 Garching, Germany}
\emailAdd{dcorradi@mpp.mpg.de}
\emailAdd{henn@mpp.mpg.de}

\abstract{We ask to what extent soft and collinear factorization determines the loop integrand of planar $\mathcal{N}=4$ super Yang--Mills theory. As a loop momentum becomes collinear with an external momentum, collinear factorization implies that the residue of the $L$-loop integrand is a universal factor times the $(L-1)$-loop integrand in the same helicity sector. Following unpublished work of Caron-Huot and one of the authors, we give an on-shell derivation of this condition. We combine it with the soft and collinear limits of the external momenta, including those that relate different helicity sectors, and impose these conditions on a general ansatz of planar, dual conformal integrands. For $n$ external legs and $L$ loops, the conditions determine the MHV and NMHV integrands uniquely in all cases with $n+L\le9$, as well as all MHV integrands with $n>4$ and $n+L=10$, and the five-point five-loop, the six-point four-loop and the nine-point one-loop NMHV integrands. The results include new local forms for the six-point four-loop MHV and NMHV integrands and the six-point three-loop NMHV integrand. All integrands agree with a BCFW-based recursion and are provided in machine-readable form.}

\preprint{MPP-2026-226}
\begin{document}
\maketitle

\newpage

\section{Introduction}
\label{sec:intro}

Explicit results for scattering amplitudes, computed at ever higher loop order and multiplicity, have repeatedly revealed structure that is hidden in the Feynman-diagram expansion. On-shell recursion relations, soft and collinear theorems, hidden symmetries and dualities, geometric formulations, and cluster-algebraic structures all emerged in this way, with planar $\mathcal{N}=4$ super Yang--Mills (sYM) theory a particularly fertile source~\cite{Arkani-Hamed:2022rwr}.

The object of these studies is usually the integrated amplitude, expressed through special functions such as multiple polylogarithms. It has become clear, however, that the loop integrand already carries much of this structure and merits study in its own right: the analytic structure of the integrated result is encoded in the singularities of the integrand. Two developments turned this into practical technology: the symbol and coproduct of transcendental functions~\cite{Goncharov:2010jf, Duhr:2012fh}, which organize polylogarithmic amplitudes, and the method of canonical differential equations~\cite{Henn:2013pwa}, which exploits the singularity structure of the integrand and is now standard in precision QCD~\cite{Henn:2020omi}.

In planar $\mathcal{N}=4$ sYM the loop integrand is a single rational function, and it is natural to ask what data determine it. Generalized unitarity matches a basis of integrals to a spanning set of cuts or leading singularities~\cite{Bern:1994zx,Arkani-Hamed:2010pyv}, sharpened at the integrand level by prescriptive unitarity~\cite{Bourjaily:2017wjl}. The planar integrand can be built recursively~\cite{Arkani-Hamed:2010zjl} from the loop-level Britto--Cachazo--Feng--Witten (BCFW) relations~\cite{Britto:2005fq}. The Amplituhedron identifies the integrand with a canonical form~\cite{Arkani-Hamed:2013jha}. A hidden permutation symmetry~\cite{Eden:2011we} has determined the four-point integrand up to twelve loops~\cite{Bourjaily:2015bpz,Bourjaily:2016evz,Bourjaily:2025iad}, through a graphical bootstrap of correlators and amplitudes~\cite{Bourjaily:2025iad,He:2024cusp}.

The logarithm of a planar amplitude has softer infrared singularities than the amplitude itself, owing to the exponentiation of soft and collinear divergences~{\cite{Sterman:2002qn,Bern:2005iz,Dixon:2008gr}}. This is already present in the integrand~\cite{Arkani-Hamed:2010pyv,Drummond:2010mb}: the leading collinear terms of the integrand of the logarithm cancel as a loop momentum becomes collinear with an external leg. In their soft-collinear bootstrap, Bourjaily, DiRe, Shaikh, Spradlin and Volovich~\cite{Bourjaily:2011hi} found that this condition, together with dual conformal invariance~\cite{Drummond:2006rz,Drummond:2007aua,Alday:2007hr} and dihedral symmetry, fixes the four-point integrand through seven loops. The same integrand is accessible from correlation functions of protected operators, which reduce in a null limit to the light-like Wilson loops dual to the amplitude~\cite{Alday:2010zy,Eden:2010zz}; there, the counterpart of the collinear condition is the operator product expansion.

Ref.~\cite{Golden:2012hi} recast the condition as a relation on the amplitude integrand rather than on its logarithm, and verified it for the known two- and three-loop MHV (maximally helicity violating) integrands with $n>4$ legs. There, the authors supplemented the condition with an additional constraint on the limit where an external momentum becomes soft~\cite{Drummond:2010mb}. Within restricted ans\"atze (based on the known form of integrands of ref.~\cite{Arkani-Hamed:2010pyv}), the two conditions together fixed the integrand at every multiplicity checked and were conjectured to do so for all $n$.

{The origin of this condition is collinear factorization. The infrared singularities of massless gauge-theory amplitudes arise from regions of loop-momentum space where virtual momenta are soft or collinear with an external momentum, and in these regions amplitudes factorize into a hard part and universal jet and soft functions, see ref.~\cite{Collins:1989gx} for a review. An all-order proof at leading power was given by Feige and Schwartz~\cite{Feige:2013zla,Feige:2014wja}. Their proof proceeds graph by graph in physical gauges, by power counting of the Feynman integrand in each soft or collinear region, with the loop momenta routed such that the region is described by one of the loop momenta. In the planar limit the dual coordinates provide such a routing for all graphs at once, so that the analysis applies to the integrand as a whole. For a single loop momentum collinear with an external momentum, it implies that the leading behavior of the $L$-loop integrand is a universal factor times the $(L-1)$-loop integrand, which is the content of the collinear condition. We thank S.~Caron-Huot for emphasizing this point to us. Beyond the planar limit there is no common labeling of the loop momenta, and making the factorization of infrared singularities manifest in the integrand requires additional care~\cite{Anastasiou:2020sdt,Anastasiou:2022eym}.}

{In planar $\mathcal{N}=4$ sYM the integrand is a single rational function of the dual coordinates, which factorizes on its cuts~\cite{Arkani-Hamed:2010zjl,Arkani-Hamed:2010pyv,Caron-Huot:2010ryg}. Caron-Huot and one of the present authors, in unpublished work~\cite{CHH}, suggested a short derivation of the collinear condition from this property, which is the starting point of this paper. It applies directly to this integrand, without reference to Feynman graphs or to a choice of gauge, and gives the universal factor in closed form in every helicity sector.}

In the present work, we establish the following:
\begin{itemize}
    \item {We give an on-shell derivation of the property that integrands satisfy when a loop momentum becomes collinear with the momentum of an external leg, a consequence of collinear factorization. The derivation gives the explicit form of this property in all helicity sectors.}
    \item   We construct {the} space of planar dual conformal integrands (with a few key assumptions on their form discussed below) and perform a bootstrap procedure using it as our ansatz. This extends the work of ref.~\cite{Golden:2012hi} to a more general space of possible integrands.
    \item We show that complementing the condition with constraints on limits in the external kinematics determines NMHV as well as MHV integrands uniquely in the cases we consider, reaching up to five loops.
\end{itemize}

\noindent For the tree-normalized integrand $I_{n,k}^{(L)}$, as a loop point $x_{A_1}$ approaches the line through two adjacent external points, $x_{A_1}\to x_1+z\,(x_2-x_1)$, {the condition reads}
\begin{equation}
\begin{aligned}
&I_{n,k}^{(L)}(x_1,\dots,x_n;\,x_{A_1},x_{A_2},\dots,x_{A_L})\;\rightarrow-\frac{I_{n,k}^{(L-1)}(x_1,\dots,x_n;\,x_{A_2},\dots,x_{A_L})}{z(z-1)\,x_{1A_1}^2\,x_{2A_1}^2}\;\,,
\end{aligned}
\label{eq:intro-condition}
\end{equation}
where {$x_{iA}=x_i-x_A$}, the $x_1,\dots,x_n$ are the external dual points and $x_{A_1},\dots,x_{A_L}$ the loop points.
We call this relation the \textit{internal collinear condition}.
Eq.~\eqref{eq:intro-condition} is equivalent to the observation of ref.~\cite{Bourjaily:2011hi} that the integrand of the logarithm of the amplitude has no collinear double pole. Carried out for the superamplitude, the derivation holds in every helicity sector, so that eq.~\eqref{eq:intro-condition} constrains integrands in MHV, NMHV and all other sectors alike. The same factorization has a first-principles interpretation in the Wilson-loop description of the integrand, where a collinear pole corresponds to a Lagrangian insertion point becoming null-separated from a cusp~\cite{Caron-Huot:2010ryg}.

To the internal collinear condition we add the soft and collinear limits of the external kinematics, under which the tree-normalized integrand reduces to the one with fewer legs~\cite{Drummond:2010mb}, as long used for one-loop amplitudes~\cite{Bern:1994zx}. Each limit has a helicity-preserving and a helicity-changing channel; the latter relates the NMHV integrand to the MHV one. The subleading soft theorem~\cite{Bern:2014oka,Cachazo:2014dia,Bianchi:2014gla} gives a further, homogeneous condition, which we use as a check.

We impose these conditions on the space of local integrands at each multiplicity and loop order. This space is built from dual conformal planar graphs, in which by assumption every propagator appears at most once, with numerators that are polynomials in momentum twistor four-brackets of complete loop lines, with rational coefficients; in the NMHV sector each term is multiplied by a dual superconformal invariant~\cite{Drummond:2008vq}.

We find that dual conformal symmetry, planarity and the infrared conditions above determine the MHV and NMHV integrands of planar $\mathcal{N}=4$ sYM uniquely for $n$ legs and $L$ loops with $n+L\le9$, as well as all MHV integrands with $n>4$ and $n+L=10$, and the five-point five-loop, the six-point four-loop and the nine-point one-loop NMHV integrands. In the four-point six-loop MHV integrand, instead, ten terms remain undetermined. Already at five loops, three terms of the four-point integrand are fixed only by the five-point conditions, in which it enters through the soft and collinear limits in external kinematics; we therefore expect that applying the procedure to the five-point integrands at six loops may determine all terms uniquely also in this case.

Our construction reproduces the known local representations and yields new ones: the six-point four-loop MHV and NMHV integrands, and the six-point three-loop NMHV integrand, for which, to the best of our knowledge, no earlier local representation had been presented. All integrands are checked to agree with the results of loop-level BCFW recursion and are provided in machine-readable form.

Section~\ref{sec:integrand} reviews the planar integrand in dual variables and superspace. Section~\ref{sec:condition} derives the internal collinear condition~\eqref{eq:intro-condition}. Section~\ref{sec:bootstrap} describes the construction of the ansatz and the application of the conditions; section~\ref{sec:results} describes the implementation and its performance, the results and the checks. Appendix~\ref{app:split} collects the tree-level splitting amplitudes used in the derivation, appendix~\ref{sec:external} reviews the conditions used for limits in external kinematics, and appendix~\ref{app:twistorform} recalls momentum twistors and gives the form of the integrand and of the conditions in these variables.

\section{Loop integrands in planar \texorpdfstring{$\mathcal{N}=4$}{N=4} sYM}
\label{sec:integrand}

In this section, we review the definition of the planar integrand in $\mathcal{N}=4$ sYM, focusing on the MHV and NMHV sectors.

\subsection{The bosonic integrand in dual variables}
\label{sec:dual}

A loop amplitude is an integrated quantity, and in general the integrand that produces it is not unique. Before formulating conditions on ``the'' integrand, let us be explicit about which ambiguities are present and how we deal with them.

The first issue is how to label the loop momenta. Each Feynman diagram comes with its own integrand, but adding the diagrams under a common integral sign requires a choice of loop momenta in each of them, and shifts of the loop momenta change the integrand without changing the integral. For planar diagrams there is a natural choice, fixed by the cyclic ordering of the external legs. We introduce dual coordinates $x_i$ via
\begin{equation}
p_i \;=\; x_{i+1}-x_i\,,\qquad x_{n+1}\equiv x_1\,,
\label{eq:dualcoords}
\end{equation}
so that momentum conservation is automatic, and associate a dual point to every face of a planar graph. The external faces are labeled by points $x_i$, and the $L$ internal faces by integration points $x_{A_1},\dots,x_{A_L}$, which play the role of the loop momenta. In these variables all planar diagrams are written in terms of the same integration variables, and the propagators are the dual distances
\begin{equation}
x_{ab}^2 \;=\; (x_a-x_b)^2\,.
\label{eq:distances}
\end{equation}
The only remaining freedom is the relabeling of the $x_{A_l}$ among themselves, which is removed by symmetrizing over them. In this way the planar integrand becomes a single rational function of the dual points.

The second issue is a genuine ambiguity. One can add to the integrand any term whose integral vanishes, for example a total derivative in the loop momenta, without changing the amplitude. Moreover, the integrated amplitude is infrared divergent and requires a regulator, and this enlarges the freedom. In dimensional regularization, scaleless integrals vanish, so that scaleless terms may be added to the integrand. Conversely, the $D$-dimensional integrand may contain terms that vanish in four dimensions but whose integrals contribute at $\mathcal{O}(\epsilon^0)$; beyond one loop, such terms appear in amplitudes computed in dimensional regularization~\cite{Bern:2008ap}.

In planar $\mathcal{N}=4$ sYM we can avoid these issues by using a definition of the integrand, as we explain presently. It is provided by the duality between amplitudes and light-like polygonal Wilson loops~\cite{Alday:2007hr,Drummond:2007aua,Brandhuber:2007yx}, in its integrand-level form in terms of Lagrangian insertions~\cite{Alday:2010zy,Eden:2010zz,Caron-Huot:2010ryg}. This gives the integrand $I_n^{(L)}$ as a correlation function of local operators at fixed, generic points $x_{A_l}$. It is a well-defined, finite rational function of four-dimensional kinematics, and it is automatically symmetric in the insertion points.

Consider the $L$-loop contribution to the tree-normalized MHV amplitude $M_n^{(L)}$,
\begin{equation}
A_n=A_n^{\text{MHV,tree}} \, M_n,\,\qquad M_n=\sum_L\left(\frac{g^2N_c}{8\pi^2}\right)^LM_n^{(L)},
\end{equation}
with the tree-level MHV amplitude
\begin{equation}
\begin{aligned}
&A_n^{\text{MHV,tree}}(1^+,\dots,i^-,\dots,j^-,\dots,n^+)=\frac{\langle ij\rangle^4}{\langle 12\rangle\langle 23\rangle\cdots\langle n1\rangle}.
  \end{aligned}
  \label{eq:M-n}
\end{equation}
The integrand $I^{(L)}_n$ is then related to it by
\begin{equation}
M_n^{(L)} \;=\; \Big(-\frac{1}{2}\Big)^{L}\,\frac{1}{L!}\int\left(\prod_{l=1}^{L}\frac{d^D x_{A_l}}{i\pi^2}\right)\;I_n^{(L)}\big(x_1,\dots,x_n;\,x_{A_1},\dots,x_{A_L}\big)\,.
\label{eq:defint}
\end{equation}
The integrand in the planar theory is obtained from the light-like polygonal Wilson loop, whose cusps are the dual points $x_i$ of the external momenta. Taking derivatives of the Wilson loop with respect to the coupling inserts the Lagrangian of the theory, integrated over spacetime. The $L$-loop integrand is then the Born-level correlation function of the Wilson loop with $L$ Lagrangian insertions, with the insertion points playing the role of the loop integration points $x_{A_1},\dots,x_{A_L}$. With the insertion points kept at generic, non-coincident positions in four dimensions, this correlation function is finite and unambiguous. It is a rational function of the dual points and is automatically symmetric under permutations of the insertion points. Equivalently, it can be obtained from correlation functions of half-BPS operators with Lagrangian insertions, in the limit where the operators become sequentially light-like separated~\cite{Alday:2010zy,Eden:2010zz}. For four points, this representation was used in ref.~\cite{Henn:2019swt}, where the explicit relation between the Born-level correlator and the integrand can be found. In the MHV sector the relevant object is the bosonic Wilson loop. In the NMHV sector it is replaced by its supersymmetric extension in twistor space~\cite{Mason:2010yk,Caron-Huot:2010ryg}, or equivalently by the light-like limit of the super-correlator~\cite{Eden:2011yp,Eden:2011ku}. At one loop, this gives the familiar box integrand for four points.

Note that we have written the integral in eq.~\eqref{eq:defint} in $D$ dimensions.
In the present paper, we study the loop integrand in four dimensions.
In practice, in the following we do not compute the integrand directly from the Wilson loop with Lagrangian insertions (or from an amplitude definition), but follow a bootstrap approach.
We then ask whether, given a space of rational functions, an integrand exists that satisfies a set of physical consistency conditions.
Concretely, we write
\begin{equation}
I_n^{(L)} \;=\; \frac{N_n^{(L)}\big(x_1,\dots,x_n;\,x_{A_1},\dots,x_{A_L}\big)}{\Big(\prod_{j=1}^{L}\prod_{i=1}^{n} x_{iA_j}^2\Big)\Big(\prod_{1\le r<l\le L} x_{A_rA_l}^2\Big)}\,,
\label{eq:numerator}
\end{equation}
which defines the rational integrand in terms of an unknown numerator $N_n^{(L)}$.
We then formulate conditions that the latter has to satisfy, and study whether a solution exists, and whether it is unique.
We leave the interesting question of whether similar criteria can be used to determine a $D$-dimensional integrand to future work.

\subsection{Symmetries}
\label{sec:symmetries}

The planar integrand is dual conformal invariant (DCI)~\cite{Drummond:2006rz,Drummond:2007aua,Alday:2007hr}. Under an inversion $x^\mu\to x^\mu/x^2$ one has
\begin{equation}
x_{ij}^2 \;\longrightarrow\; \frac{x_{ij}^2}{x_i^2\,x_j^2}\,,\qquad d^4x_A \;\longrightarrow\; \frac{d^4x_A}{(x_A^2)^4}\,,
\label{eq:inversion}
\end{equation}
so that invariance of eq.~\eqref{eq:defint} requires the integrand to carry uniform conformal weight four at each loop point and weight zero at each external point. Since every propagator in the denominator of eq.~\eqref{eq:numerator} has weight one at each of its two points, this fixes the weight of the numerator $N_n^{(L)}$ completely.

A simple example is the one-loop four-point integrand
\begin{equation}
I_4^{(1)} \;=\; \frac{x_{13}^2\,x_{24}^2}{x_{1A}^2\,x_{2A}^2\,x_{3A}^2\,x_{4A}^2}\,.
\label{eq:box}
\end{equation}
Here $x_{13}^2x_{24}^2$ is exactly what is needed to balance the weights of the four propagators.

In $\mathcal{N}=4$ sYM, the states of the multiplet are collected into a single on-shell superfield by means of Grassmann variables $\eta_i^A$, $A=1,\dots,4$, in the fundamental representation of the $SU(4)$ R-symmetry of the theory~\cite{Nair:1988bq}. Amplitudes for all processes of some multiplicity $n$ are then packaged into a superamplitude $\mathcal{A}_n(\lambda_i,\tilde\lambda_i,\eta_i)$. Expanding it in the Grassmann degree in $\eta_i$ separates it into sectors, whose components encode the different helicity configurations of the external states, related to each other by supersymmetric Ward identities. The term of Grassmann degree $8+4k$ is the N$^k$MHV sector; its pure-gluon component has $k+2$ negative-helicity gluons, and the sector with $k=0$ is MHV. The tree-level MHV superamplitude is the Parke--Taylor factor~\cite{Parke:1986gb},
\begin{equation}
\mathcal{A}_n^{\text{MHV,tree}} \;=\; \frac{\delta^4\big(\sum_i p_i\big)\,\delta^{0|8}\big(\sum_i\lambda_i\eta_i\big)}{\langle12\rangle\langle23\rangle\cdots\langle n1\rangle}\,.
\label{eq:PT}
\end{equation}
Let $\mathcal{M}_{n,k}$ be the N$^k$MHV amplitude, normalized by this tree-level expression,
\begin{equation}
\mathcal{A}_{n,k}=\mathcal{A}_n^{\text{MHV,tree}}\mathcal{M}_{n,k}\,.
\end{equation}
We write the expansion in the coupling as
\begin{equation}
\mathcal{M}_{n,k}=\sum_L\left(\frac{g^2N_c}{8\pi^2}\right)^L\mathcal{M}_{n,k}^{(L)} \,.
\end{equation}
Moreover, let $I_{n,k}^{(L)}$ be the integrand of $\mathcal{M}_{n,k}^{(L)}$, in analogy to eq.~\eqref{eq:defint}, generalizing this equation to any helicity. The bosonic MHV integrand $I_{n}^{(L)}$ corresponds in this notation to $I_{n,0}^{(L)}$.

The dual variables extend to superspace. Alongside eq.~\eqref{eq:dualcoords}, Grassmann-odd dual coordinates $\theta_i$ are defined as
\begin{equation}
\lambda_i\,\eta_i \;=\; \theta_{i+1}-\theta_i\,,
\label{eq:dualtheta}
\end{equation}
so that conservation of supermomentum is automatic in the same way that momentum conservation is. In these variables, the theory possesses a dual superconformal symmetry, whose bosonic part is the dual conformal symmetry discussed above~\cite{Drummond:2008vq}.

The dependence of an N$^k$MHV superamplitude on the Grassmann variables can be expanded {in} a basis of dual superconformal invariants of Grassmann degree $4k$~\cite{Drummond:2008vq}, called $R$-invariants. For the NMHV sector, these are
\begin{equation}
\begin{aligned}
R_{r;st} &= \frac{\langle s\,s{-}1\rangle\langle t\,t{-}1\rangle\;\delta^{0|4}\big(\langle r|x_{rs}x_{st}|\theta_{tr}\rangle+\langle r|x_{rt}x_{ts}|\theta_{sr}\rangle\big)}{x_{st}^2\,\langle r|x_{rs}x_{st}|t\rangle\,\langle r|x_{rs}x_{st}|t{-}1\rangle\,\langle r|x_{rt}x_{ts}|s\rangle\,\langle r|x_{rt}x_{ts}|s{-}1\rangle}\,,\\
\end{aligned}
\label{eq:Rinv}
\end{equation}
with $x_{ij}=x_i-x_j$, $\theta_{ij}=\theta_i-\theta_j$ and $\langle r|x_{rs}x_{st}|t\rangle=\lambda_r^\alpha(x_{rs})_{\alpha\dot\alpha}(x_{st})^{\dot\alpha\beta}\lambda_{t\,\beta}$.

For the loop integrand, this fixes the shape of the NMHV ansatz. Denoting the $R$-invariants by $R^{k}_a$, the N$^k$MHV super-integrand takes the form
\begin{equation}
I_{n,k}^{(L)} \;=\; \frac{\sum_a R_a^{k}\big(x_i,\theta_i\big)\,N_{a,k}^{(L)}\big(x_1,\dots,x_n;\,x_{A_1},\dots,x_{A_L}\big)}{\Big(\prod_{j=1}^{L}\prod_{i=1}^{n} x_{iA_j}^2\Big)\Big(\prod_{1\le r<l\le L} x_{A_rA_l}^2\Big)}\,.
\label{eq:nmhvdual}
\end{equation}
Each $N_{a,k}^{(L)}$ carries the same dual conformal weight as the corresponding MHV numerator $N_n^{(L)}$, in order for the whole object to be DCI. The scope of the present paper is $k\leq1$: MHV and NMHV integrands.

\section{The internal collinear condition}
\label{sec:condition}
In this section we derive the internal collinear condition~\eqref{eq:intro-condition} as a universal property of planar integrands of $\mathcal{N}=4$ sYM. Following unpublished work of S.~Caron-Huot and one of the authors~\cite{CHH}, the derivation rests on the factorization properties of the integrand. Let us outline the main steps.
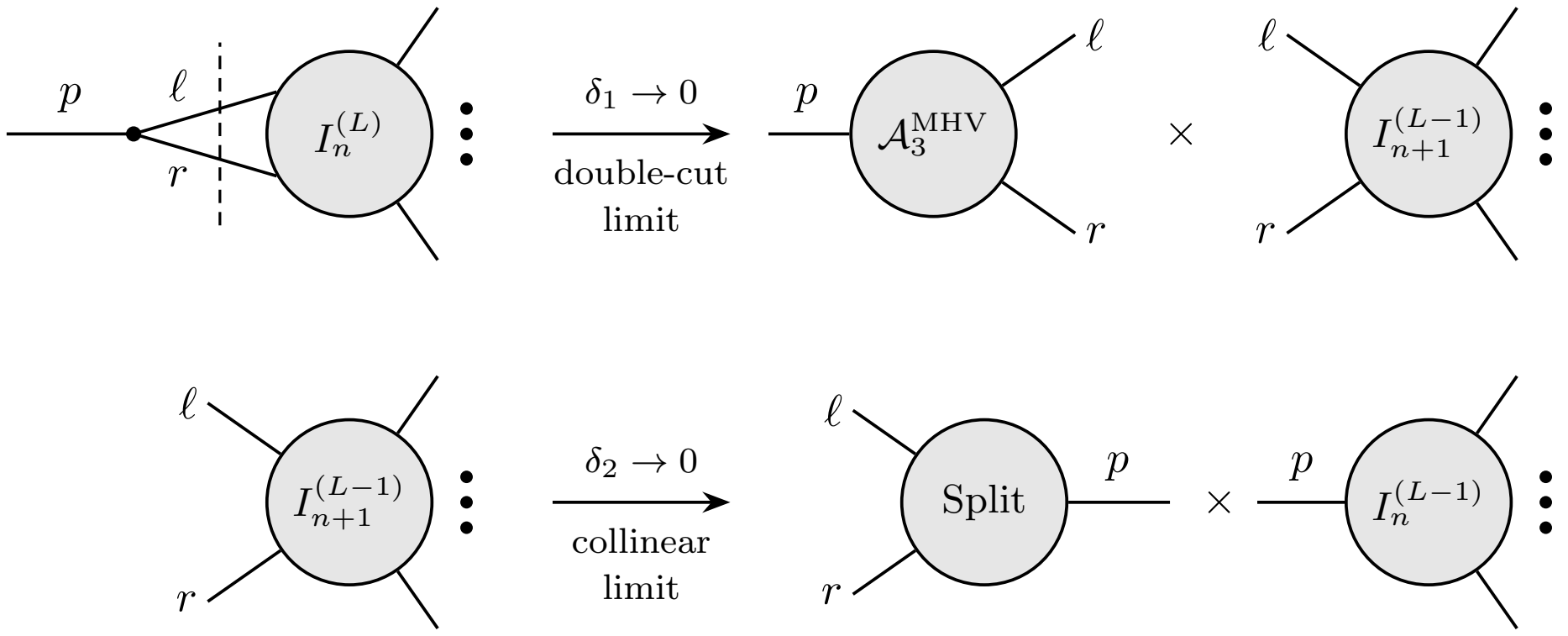
\begin{figure}[t]
\centering
\begin{tikzpicture}[line width=0.7pt, every node/.style={font=\small},
  blob/.style={circle, draw, fill=black!10, minimum size=1.3cm, inner sep=1pt, font=\footnotesize},
  lim/.style={-{Stealth[length=2.2mm]}},
  cut/.style={densely dashed, line width=0.6pt}]
\node[blob] (L) at (0.9,0) {$ I_n^{(L)}$};
\draw (-1.8,0) -- (-0.8,0); \node[above] at (-1.3,0.03) {$p$};
\fill (-0.8,0) circle (1.7pt);
\draw (-0.8,0) -- (L.150); \node[above] at (-0.45,0.12) {$\ell$};
\draw (-0.8,0) -- (L.210); \node[below] at (-0.45,-0.12) {$r$};
\draw[cut] (-0.12,-0.72) -- (-0.12,0.72);
\draw (L.55) -- ++(55:0.55); \draw (L.-55) -- ++(-55:0.55); \foreach \dy in {-0.2,0,0.2} \fill ($(L.east)+(0.26,\dy)$) circle (1.4pt);
\draw[lim] (2.5,0) -- (3.9,0);
\node[above, font=\scriptsize] at (3.2,0.06) {$\delta_1\to0$};
\node[below, font=\scriptsize, align=center] at (3.2,-0.06) {double-cut\\ limit};
\node[blob] (A) at (5.5,0) {$\mathcal{A}_3^{\text{MHV}}$};
\draw (4.2,0) -- (A.180); \node[above] at (4.5,0.03) {$p$};
\draw (A.35) -- ++(35:0.7) node[right, inner sep=2pt] {$\ell$};
\draw (A.-35) -- ++(-35:0.7) node[right, inner sep=2pt] {$r$};
\node at (7.45,0) {$\times$};
\node[blob] (B) at (9.4,0) {$ I_{n+1}^{(L-1)}$};
\draw (B.145) -- ++(145:0.7) node[left, inner sep=2pt] {$\ell$};
\draw (B.215) -- ++(215:0.7) node[left, inner sep=2pt] {$r$};
\draw (B.55) -- ++(55:0.55); \draw (B.-55) -- ++(-55:0.55); \foreach \dy in {-0.2,0,0.2} \fill ($(B.east)+(0.26,\dy)$) circle (1.4pt);
\def\yb{-2.9}
\node[blob] (C) at (0.9,\yb) {$ I_{n+1}^{(L-1)}$};
\draw (C.145) -- ++(145:0.7) node[left, inner sep=2pt] {$\ell$};
\draw (C.215) -- ++(215:0.7) node[left, inner sep=2pt] {$r$};
\draw (C.55) -- ++(55:0.55); \draw (C.-55) -- ++(-55:0.55); \foreach \dy in {-0.2,0,0.2} \fill ($(C.east)+(0.26,\dy)$) circle (1.4pt);
\draw[lim] (2.5,\yb) -- (3.9,\yb);
\node[above, font=\scriptsize] at (3.2,\yb+0.06) {$\delta_2\to0$};
\node[below, font=\scriptsize, align=center] at (3.2,\yb-0.06) {collinear\\ limit};
\node[blob] (S) at (5.9,\yb) {Split};
\draw (S.145) -- ++(145:0.6) node[left, inner sep=2pt] {$\ell$};
\draw (S.215) -- ++(215:0.6) node[left, inner sep=2pt] {$r$};
\draw (S.0) -- ++(0:0.8); \node[above] at (6.95,\yb+0.03) {$p$};
\node at (7.75,\yb) {$\times$};
\node[blob] (D) at (9.4,\yb) {$ I_n^{(L-1)}$};
\draw (8.05,\yb) -- (D.180); \node[above] at (8.4,\yb+0.03) {$p$};
\draw (D.55) -- ++(55:0.55); \draw (D.-55) -- ++(-55:0.55); \foreach \dy in {-0.2,0,0.2} \fill ($(D.east)+(0.26,\dy)$) circle (1.4pt);
\end{tikzpicture}
\caption{Internal collinear limit. Top: in the double-cut limit, the $L$-loop integrand with the propagators $\ell$ and $r=p-\ell$ adjacent to the external leg $p$ factorizes into a three-point MHV amplitude times an $(L-1)$-loop integrand with legs $\ell$ and $r$, summed over their states. Bottom: in the collinear limit, $\ell$ and $r$ merge into $p$, and the $(L-1)$-loop integrand with the legs $\ell$ and $r$ factorizes into a splitting amplitude times an $(L-1)$-loop integrand.}
\label{fig:intcoll}
\end{figure}
\begin{itemize}
    \item We consider the pole of the integrand in the limit where a loop momentum becomes collinear with an external one. As depicted in the top half of figure~\ref{fig:intcoll}, the integrand factorizes into the product of terms on the two sides of the cut: a (tree-level) three-point amplitude and an $(L-1)$-loop integrand of $(n+1)$ particles, with an integration over Grassmann variables corresponding to states crossing the cut propagators (not shown in the figure).
    \item The $(L-1)$-loop, $(n+1)$-point integrand resulting from the cut is itself in a collinear configuration; therefore it reduces to a product of a three-point amplitude (the splitting amplitude), and the $(L-1)$-loop, $n$-point integrand, again with an integration over the Grassmann variable crossing the cut. This is depicted in the bottom half of figure~\ref{fig:intcoll}.
    \item The weight in Grassmann variables of each term enforces selection rules in the factorization. After writing the factorized expression, terms that shift the helicity sector of integrands cancel. Performing Grassmann integrations, we recover expression~\eqref{eq:intro-condition}.
\end{itemize}
\subsection{Example: the two-loop, four-point integrand}
\begin{figure}[t]
\centering
\begin{tikzpicture}[x=1cm,y=1cm, prop/.style={line width=0.8pt}, cut/.style={densely dotted, line width=1.1pt},
  lab/.style={font=\scriptsize}, num/.style={font=\scriptsize}]
\begin{scope}[yshift=0.6cm]
\draw[prop] (0,0) rectangle (2.4,1.2); \draw[prop] (1.2,0) -- (1.2,1.2);
\draw[prop] (0,1.2) -- (-0.35,1.55) node[num, above left=-2pt] {1}; \draw[prop] (2.4,1.2) -- (2.75,1.55) node[num, above right=-2pt] {2};
\draw[prop] (2.4,0) -- (2.75,-0.35) node[num, below right=-2pt] {3}; \draw[prop] (0,0) -- (-0.35,-0.35) node[num, below left=-2pt] {4};
\node[lab] at (0.6,0.6) {$A$}; \node[lab] at (1.8,0.6) {$B$};
\draw[cut] (0.6,0.92) -- (0.6,1.48); \draw[cut] (-0.3,0.6) -- (0.2,0.6);
\node[lab] at (1.2,-1.5) {(a)};
\end{scope}
\begin{scope}[xshift=4.1cm, yshift=0.6cm]
\draw[prop] (0,0) rectangle (2.4,1.2); \draw[prop] (1.2,0) -- (1.2,1.2);
\draw[prop] (0,1.2) -- (-0.35,1.55) node[num, above left=-2pt] {1}; \draw[prop] (2.4,1.2) -- (2.75,1.55) node[num, above right=-2pt] {2};
\draw[prop] (2.4,0) -- (2.75,-0.35) node[num, below right=-2pt] {3}; \draw[prop] (0,0) -- (-0.35,-0.35) node[num, below left=-2pt] {4};
\node[lab] at (0.6,0.6) {$B$}; \node[lab] at (1.8,0.6) {$A$};
\draw[cut] (1.8,0.92) -- (1.8,1.48);
\node[lab] at (1.2,-1.5) {(b)};
\end{scope}
\begin{scope}[xshift=8.4cm]
\draw[prop] (0,0) rectangle (1.2,2.4); \draw[prop] (0,1.2) -- (1.2,1.2);
\draw[prop] (0,2.4) -- (-0.35,2.75) node[num, above left=-2pt] {1}; \draw[prop] (1.2,2.4) -- (1.55,2.75) node[num, above right=-2pt] {2};
\draw[prop] (1.2,0) -- (1.55,-0.35) node[num, below right=-2pt] {3}; \draw[prop] (0,0) -- (-0.35,-0.35) node[num, below left=-2pt] {4};
\node[lab] at (0.6,1.8) {$A$}; \node[lab] at (0.6,0.6) {$B$};
\draw[cut] (0.6,2.12) -- (0.6,2.68); \draw[cut] (-0.3,1.8) -- (0.2,1.8);
\node[lab] at (0.6,-0.9) {(c)};
\end{scope}
\begin{scope}[xshift=11.3cm]
\draw[prop] (0,0) rectangle (1.2,2.4); \draw[prop] (0,1.2) -- (1.2,1.2);
\draw[prop] (0,2.4) -- (-0.35,2.75) node[num, above left=-2pt] {1}; \draw[prop] (1.2,2.4) -- (1.55,2.75) node[num, above right=-2pt] {2};
\draw[prop] (1.2,0) -- (1.55,-0.35) node[num, below right=-2pt] {3}; \draw[prop] (0,0) -- (-0.35,-0.35) node[num, below left=-2pt] {4};
\node[lab] at (0.6,1.8) {$B$}; \node[lab] at (0.6,0.6) {$A$};
\draw[cut] (-0.3,0.6) -- (0.2,0.6);
\node[lab] at (0.6,-0.9) {(d)};
\end{scope}
\end{tikzpicture}
\caption{The four terms in $I^{(2)}_4$, drawn as double boxes in momentum space with the external legs $1,\dots,4$; the labels $A$, $B$ mark the loop points $x_{A}$, $x_B$. The dotted lines indicate the propagators $x_{1A}^2$ and $x_{2A}^2$, which go on shell on the locus of the collinear condition.
}
\label{fig:dboxes}
\end{figure}
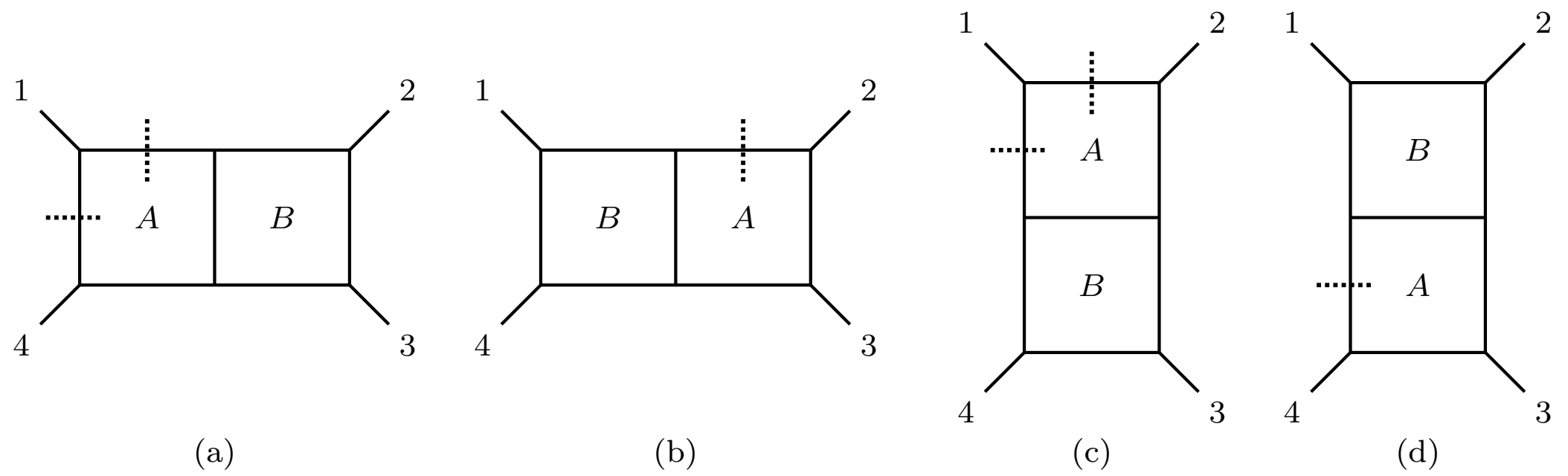

Before we discuss the details of the derivation outlined above, it is useful to familiarize ourselves with the condition by considering the example of the two-loop, four-point integrand. This is given by~{\cite{Bern:1997nh}}
\begin{equation}
\begin{aligned}
I^{(2)}_{4,0}=&\frac{x_{13}^2\,(x_{24}^2)^2}{x_{1A}^2x_{2A}^2x_{4A}^2\,x_{AB}^2\,x_{2B}^2x_{3B}^2x_{4B}^2}\,+\frac{x_{13}^2\,(x_{24}^2)^2}{x_{1B}^2x_{2B}^2x_{4B}^2\,x_{AB}^2\,x_{2A}^2x_{3A}^2x_{4A}^2}\\
&+\frac{x_{24}^2\,(x_{13}^2)^2}{x_{1A}^2x_{2A}^2x_{3A}^2\,x_{AB}^2\,x_{3B}^2x_{4B}^2x_{1B}^2}\,+\frac{x_{24}^2\,(x_{13}^2)^2}{x_{1B}^2x_{2B}^2x_{3B}^2\,x_{AB}^2\,x_{3A}^2x_{4A}^2x_{1A}^2}\,.
\end{aligned}
\label{eq:dboxterm}
\end{equation}
 Its four terms are represented in figure~\ref{fig:dboxes}.

We let $x_A$ approach the collinear region $(1-z) x_1+zx_2$, in which the propagators $x_{1A}^2$, $x_{2A}^2$ go on shell.
Only the first and third {terms} in eq.~\eqref{eq:dboxterm} contain both of these propagators (see also figure~\ref{fig:dboxes}(a), (c)).
In this limit, $x^2_{4A}=z\,x_{24}^2$, $x^2_{3A}=(1-z)\,x_{13}^2$ and $x^2_{AB}=(1-z)\,x^2_{1B}+z\,x^2_{2B}$, so that
\begin{equation}
\begin{aligned}
I^{(2)}_{4,0}&\rightarrow\frac{1}{x_{1A}^2x_{2A}^2}\frac{x_{13}^2x_{24}^2}{(1-z)x^2_{1B}+zx^2_{2B}}\left(\frac{1}{zx^2_{2B}x^2_{3B}x^2_{4B}}+\frac{1}{(1-z)x^2_{1B}x^2_{3B}x^2_{4B}}\right)\\&=-\frac{1}{x_{1A}^2x_{2A}^2}\frac{1}{z(z-1)}\frac{x_{13}^2x_{24}^2}{x^2_{1B}x^2_{2B}x^2_{3B}x^2_{4B}}.
\label{eq:dboxlocus}
\end{aligned}
\end{equation}
Comparing the last line with the one-loop integrand in eq.~\eqref{eq:box}, we find that eq.~\eqref{eq:intro-condition} is satisfied.
We see from the derivation that eq.~\eqref{eq:intro-condition} relies on a nontrivial interplay between different terms. When writing the integrand in the form of eq.~\eqref{eq:numerator}, this becomes a condition on the numerator polynomial $N_n^{(L)}$.

\subsection{Details of the derivation}
\label{sec:statement}

Consider an integrand $I_{n,k}^{(L)}$, written in loop-momentum space. Let $p$ be an external momentum, and $\ell$ and $r= p-\ell$ the momenta of the two propagators adjacent to it, as in figure~\ref{fig:intcoll}.
\begin{enumerate}
    \item \textit{Parameterization of the limit.} We parameterize $\ell$ so that its square and angle spinors become aligned with those of $p$ in two separate steps:
\begin{equation}
\ell=|\ell_l\rangle[\ell_l|+\delta_1 c\,|q\rangle[q|\,,\qquad
\begin{aligned}
|\ell_l]&=\sqrt z\,|p]+\delta_1\sqrt{1-z}\,|q]\,,\\
|\ell_l\rangle&=\sqrt z\,|p\rangle+\delta_2\sqrt{1-z}\,|q\rangle\,,
\end{aligned}
\label{eq:kpar}
\end{equation}
with an auxiliary light-like reference $q$, and $c$ a generic constant. For the other momentum, we write $r=|r_l\rangle[r_l|-\delta_1(c+\delta_2)\,|q\rangle[q|$, where the spinors of $r_l$ follow from those of $\ell_l$ by $z\to1-z$ and $\delta_{1,2}\to-\delta_{1,2}$. The parameter $\delta_1$ enters only the square spinors, $\delta_2$ only the angle ones, and
\begin{equation}
\begin{aligned}
\ell^2&=z\,\delta_1 c\,\langle pq\rangle[qp]\,,\qquad & r^2&=-(1-z)\,\delta_1(c+\delta_2)\,\langle pq\rangle[qp]\,,\\
[\ell_lr_l]&=-\delta_1[pq]\,,& \langle \ell_lr_l\rangle&=-\delta_2\langle pq\rangle\,,
\end{aligned}
\label{eq:invariants}
\end{equation}
so that $P^2\equiv(\ell_l+r_l)^2=\delta_1\delta_2\,\langle pq\rangle[qp]$. As $\delta_1\to0$ both propagators vanish. The double cut $\ell^2=r^2=0$ has two branches,
\begin{equation}
\text{MHV:}\ [\ell_lr_l]=0\,,\ \ \langle \ell_lr_l\rangle\neq0\,,\qquad\qquad
\MHVb\text{:}\ \langle \ell_lr_l\rangle=0\,,\ \ [\ell_lr_l]\neq0\,,
\label{eq:branches}
\end{equation}
on which the three-point MHV and $\MHVb$ amplitudes, respectively, have support. The intersection of these two is the real collinear point $\ell=zp$. We call $\delta_1\to0$ the double-cut limit: it takes a generic $\ell$ onto the MHV branch, with the square spinors aligned. We call the subsequent $\delta_2\to0$ the collinear limit: it aligns the angle spinors as well and moves $\ell$ along the MHV branch to the collinear point.
\item \textit{Factorization on the double cut.} In the double-cut limit $\delta_1\rightarrow0$, integrand factorization gives the residue as a three-point tree amplitude times an integrand with one leg more and one loop fewer, summed over the states that can flow through the two cut lines (figure~\ref{fig:intcoll}):
\begin{equation}
\begin{aligned}
\ell^2(p-\ell)^2\,\big(\mathcal{A}_n^{\text{MHV,tree}}\,&{I_{n,k}^{(L)}}\big)\Big[p^{I_p},\dots;\,\ell,\ell_1,\dots,\ell_{L-1}\Big]\\
\overset{\delta_1\rightarrow0}{=}&{}-\sum_{I_1,I_2}\mathcal{A}_3^{\text{MHV}}\Big[p^{I_p},-(p-\ell)^{\bar I_2},-\ell^{\bar I_1}\Big]\\
&\times\big(\mathcal{A}_{n+1}^{\text{MHV,tree}}\,{I_{n+1,k}^{(L-1)}}\big)\Big[\ell^{I_1},(p-\ell)^{I_2},\dots;\ell_1,\dots,\ell_{L-1}\Big],
\end{aligned}
\label{eq:fixedloopfact}
\end{equation}
where the three-point amplitude is the MHV one, the only one with support on the MHV branch~\eqref{eq:branches}.
The sum over the states of each cut line, the sixteen components of one superfield (section~\ref{sec:symmetries}), is an integral over the Grassmann variables of that line, which pairs each state $I_i$ with its conjugate $\bar I_i$ on the other side of the cut. In the parameterization~\eqref{eq:kpar}, the amplitude~\eqref{eq:A3super} is
\begin{equation}
\begin{aligned}
&\mathcal{A}_3^{\text{MHV}}\big[p,-\ell,\ell-p\,;\,\eta_p,\eta_\ell,\eta_{(\ell-p)}\big]\\{}=-&\frac{\delta_2\langle pq\rangle}{\sqrt{z(1-z)}}\;\delta^{0|4}\big(\sqrt z\,\eta_\ell+\sqrt{1-z}\,\eta_{(\ell-p)}-\eta_p\big)\delta^{0|4}\big(\sqrt{1-z}\,\eta_\ell-\sqrt z\,\eta_{(\ell-p)}\big)\,.
\end{aligned}
\label{eq:A3par}
\end{equation}

\item \textit{Factorization of the lower-loop integrand.} On the MHV branch the square spinors of the cut legs $\ell$ and $p-\ell$ are aligned. For the lower-loop integrand, of which they are adjacent external legs, this leads to a collinear pole. Let us write its expansion,
\begin{equation}
\begin{aligned}
\Big(\mathcal{A}_{n+1}^{\text{MHV,tree}}\,I_{n+1,k}^{(L-1)}\Big)\big[\ell,\eta_\ell;\,p-\ell,\eta_{(\ell-p)};\,\dots\big]&\\
\overset{\delta_1\rightarrow0}{=}{}\int d^4\eta'_p\;\mathrm{Split}^{\text{MHV}}\big[z;\,\eta_\ell,\eta_{(\ell-p)},\eta'_p\big]&\mathcal{A}_n^{\text{MHV,tree}}\big[p,\eta'_p;\,\dots\big]\,I_{n,k-1}^{(L-1)}\big[p,\eta'_p;\,\dots\big]\\
+\Big(\mathcal{A}_{n+1}^{\text{MHV,tree}}&\,I_{n+1,k}^{(L-1)}\Big)\big[\ell,\eta_\ell;\,p-\ell,\eta_{(\ell-p)};\,\dots\big]\Big|_{\delta_1^0}\,,
\end{aligned}
\label{eq:lowerexp}
\end{equation}
where the subscript $|_{\delta_1^0}$ represents the subleading order of the integrand, constant in $\delta_1$. The leading term is given by the MHV splitting amplitude, which has a simple pole in $\delta_1$ (see eq.~\eqref{eq:splitA3}). This contribution drops out of eq.~\eqref{eq:fixedloopfact}: the MHV splitting amplitude contains the same delta function as the one that appears in the last line of eq.~\eqref{eq:A3par}, so that the product contains it twice and vanishes by nilpotency.
\item \textit{Collinear limit and Grassmann integration.} We take the limit $\delta_2\rightarrow0$, where the subleading term in the expansion of the lower-loop integrand~\eqref{eq:lowerexp} has a pole. The residue is given by
\begin{equation}
\begin{aligned}
\Big(\mathcal{A}_{n+1}^{\text{MHV,tree}}\,I_{n+1,k}^{(L-1)}\Big)&\big[\ell,\eta_\ell;\,p-\ell,\eta_{(\ell-p)};\,\dots\big]\Big|_{\delta_1^0}\\
\overset{\delta_2\rightarrow0}{=}{}&\int d^4\eta'_p\;\mathrm{Split}^{\MHVb}\big[z;\,\eta_\ell,\eta_{(\ell-p)},\eta'_p\big]\\
&\times\mathcal{A}_n^{\text{MHV,tree}}\big[p,\eta'_p;\,\dots\big]\,I_{n,k}^{(L-1)}\big[p,\eta'_p;\,\dots\big],
\end{aligned}
\label{eq:constpole}
\end{equation}
with the $\MHVb$ splitting amplitude, which has a simple pole in $\langle \ell_lr_l\rangle\propto\delta_2$.
Inserting eq.~\eqref{eq:constpole} in eq.~\eqref{eq:fixedloopfact}, the powers of $\delta_1$ and $\delta_2$ and the brackets involving $q$, which appear in the numerators of three-point amplitudes, cancel against a factor $(\ell_l+r_l)^2$ in the denominator of the $\MHVb$ splitting amplitude. We obtain
\begin{equation}
\begin{aligned}
\lim_{\delta_2\to0}\lim_{\delta_1\to0}\;&\ell^2(p-\ell)^2\,\mathcal{A}_n^{\text{MHV,tree}}\big[p,\eta_p,\dots\big]\,I_{n,k}^{(L)}\big[p,\eta_p,\dots;\,\ell,\ell_1,\dots,\ell_{L-1}\big]\\
&=\frac{1}{z(1-z)}\int d^4\eta'_p\,d^4\eta_\ell\,d^4\eta_{(\ell-p)}\;\delta^{0|4}\big(\eta'_p-\eta_p\big)\,\\
&\times\delta^{0|4}\big(\sqrt z\,\eta_\ell+\sqrt{1-z}\,\eta_{(\ell-p)}-\eta_p\big)\;\delta^{0|4}\big(\sqrt{1-z}\,\eta_\ell-\sqrt z\,\eta_{(\ell-p)}\big)\\
&\times\;\mathcal{A}_n^{\text{MHV,tree}}\big[p,\eta'_p,\dots\big]\,I_{n,k}^{(L-1)}\big[p,\eta'_p,\dots;\,\ell_1,\dots,\ell_{L-1}\big]\,.
\end{aligned}
\label{eq:chain}
\end{equation}
Evaluating the Grassmann integral then gives eq.~\eqref{eq:intro-condition}, which completes our derivation.
\end{enumerate}
\section{Bootstrapping the integrand}
\label{sec:bootstrap}
 In this section, we outline a bootstrap procedure that allows us to determine integrands using the condition~\eqref{eq:intro-condition} derived above, together with other standard conditions and properties. The bootstrap fixes the integrand $I_{n,k}^{(L)}$ for one number of legs $n$ and one loop order $L$ at a time; we call this the case $(n,L)$. The integrand, a rational function of external kinematics and loop variables, is written as a linear combination of a set of terms with unknown rational coefficients.

To determine the coefficients, we impose the internal collinear condition of section~\ref{sec:condition} along with properties of limits in the external kinematics, collected in appendix~\ref{sec:external}. The conditions relate the integrand with $n$ legs at $L$ loops to the integrands $(n,L-1)$ and $(n-1,L)$ of the same sector; in the NMHV sector, some conditions, which we refer to as \textit{helicity-changing}, relate it to the MHV integrand $(n-1,L)$. The cases are therefore processed in a fixed order. In the MHV sector we start from $(4,1)$ and raise the loop order at fixed multiplicity up to $n+L=9$ (or $n+L=10$ in a few additional cases, as specified below); we then raise the multiplicity by one and start again from one loop. The NMHV sector is treated afterwards, in the same order. In this way every integrand that enters the conditions is known when it is needed.
\subsection{Construction of the ansatz}

\label{sec:ansatz}

We list here a set of assumptions about the integrands, on which we base our construction of ans\"atze for the bootstrap.
\begin{itemize}
\item \emph{Dual conformal invariance, planarity, and single propagator powers}. The ansatz is built as a linear combination of rational functions, each of which must be DCI. Their denominators must correspond to planar diagrams. In addition, we assume that every propagator appears at most once in the denominator of each term. This imposes strict bounds on the set of propagators and numerators that can appear. The local integrands of refs.~\cite{Arkani-Hamed:2010pyv,Bourjaily:2011hi,Bourjaily:2015jna} have this form.

\item \emph{Polynomial numerators.} We assume that the numerator of each term is a polynomial in four-brackets of complete loop lines, and the coefficients of the linear combination are rational numbers; in the NMHV sector the integrand is such a linear combination of products of terms with $R$-invariants. A term then has no poles in the external kinematics, apart from those of the $R$-invariants. For $n\le5$ the external twistors admit no dual conformal cross-ratios, so that a dual conformally invariant coefficient is necessarily a number. For $n\ge6$ a coefficient could depend on cross-ratios; we assume that it does not.
\item \emph{Dihedral symmetry.} The tree-normalized integrand is invariant under a dihedral transformation of the labeling of external momenta, generated by the cyclic shift $i\to i+1$ and the reversal of the order of the legs~\cite{Mangano:1990by}, and symmetric in the loop points; and so must be all elements of our ansatz.
\end{itemize}

\noindent These assumptions are supported, a posteriori, by the fact that results derived from these ans\"atze provide the correct integrand, equal in all cases we studied to those obtained from recursion relations, as we discuss in section~\ref{sec:checks}.
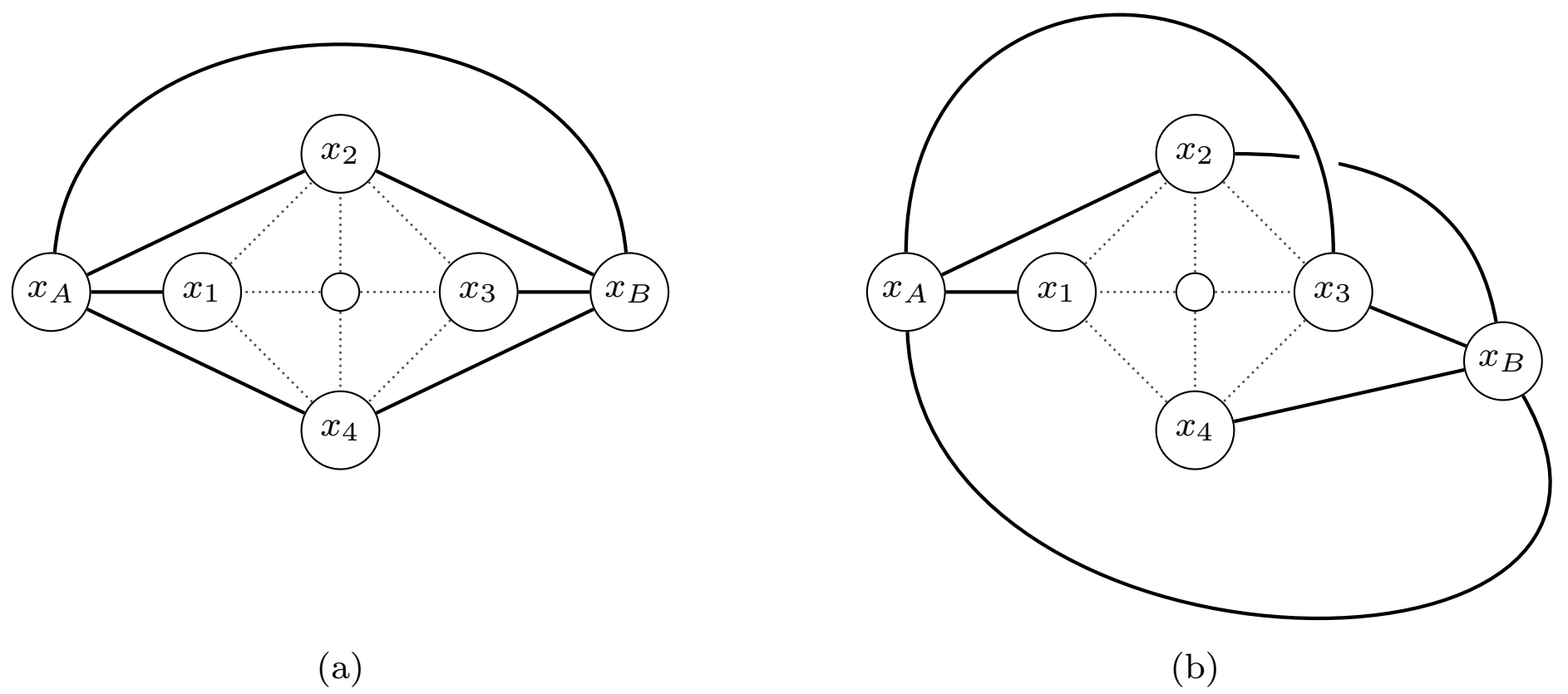
\begin{figure}[t]
\centering
\begin{tikzpicture}[x=1cm,y=1cm, prop/.style={line width=0.8pt}, aux/.style={densely dotted, line width=0.5pt, black!70},
  dual/.style={circle, draw, fill=white, inner sep=0.5pt, font=\scriptsize, minimum size=0.62cm},
  auxv/.style={circle, draw, fill=white, inner sep=0pt, minimum size=0.3cm}, lab/.style={font=\scriptsize}]
\begin{scope}
\node[auxv] (c) at (0,0) {};
\node[dual] (x2) at (0,1.1) {$x_2$}; \node[dual] (x3) at (1.1,0) {$x_3$}; \node[dual] (x4) at (0,-1.1) {$x_4$}; \node[dual] (x1) at (-1.1,0) {$x_1$};
\draw[aux] (x1) -- (x2) -- (x3) -- (x4) -- (x1); \foreach \p in {x1,x2,x3,x4} \draw[aux] (c) -- (\p);
\node[dual] (a1) at (-2.3,0) {$x_A$}; \node[dual] (a2) at (2.3,0) {$x_B$};
\draw[prop] (a1) -- (x1); \draw[prop] (a1) -- (x2); \draw[prop] (a1) -- (x4);
\draw[prop] (a2) -- (x2); \draw[prop] (a2) -- (x3); \draw[prop] (a2) -- (x4);
\draw[prop] (a1) to[out=85,in=95,looseness=1.25] (a2);
\node[lab] at (0,-3.0) {(a)};
\end{scope}
\begin{scope}[xshift=6.8cm]
\node[auxv] (c) at (0,0) {};
\node[dual] (x2) at (0,1.1) {$x_2$}; \node[dual] (x3) at (1.1,0) {$x_3$}; \node[dual] (x4) at (0,-1.1) {$x_4$}; \node[dual] (x1) at (-1.1,0) {$x_1$};
\draw[aux] (x1) -- (x2) -- (x3) -- (x4) -- (x1); \foreach \p in {x1,x2,x3,x4} \draw[aux] (c) -- (\p);
\node[dual] (a1) at (-2.3,0) {$x_A$}; \node[dual] (a2) at (2.45,-0.55) {$x_B$};
\draw[prop] (a1) -- (x1); \draw[prop] (a1) -- (x2);
\draw[prop] (a1) to[out=-88,in=-60,looseness=1.5] (a2);
\draw[prop] (a2) -- (x3); \draw[prop] (a2) -- (x4);
\path[name path=under] (x2) to[out=0,in=100,looseness=1.1] (a2);
\path[name path=over] (a1) to[out=90,in=90,looseness=1.9] (x3);
\draw[prop] (x2) to[out=0,in=100,looseness=1.1] (a2);
\fill[white, name intersections={of=under and over, by=X}] (X) circle (4.5pt);
\draw[prop] (a1) to[out=90,in=90,looseness=1.9] (x3);
\node[lab] at (0,-3.0) {(b)};
\end{scope}
\end{tikzpicture}
\caption{The planarity test at four points and two loops. (a) The dual graph of the double box; it passes all tests. (b) A graph that fails the test, although it is planar without the auxiliary edges. All dual points are drawn as circles, the small empty circle at the center represents the auxiliary vertex. Solid lines are propagators and dotted lines auxiliary edges.}
\label{fig:planarity}
\end{figure}
\noindent The terms of the ansatz are rational functions whose denominator corresponds to graphs in dual space. We construct them in the following seven steps, represented as numbered boxes in figure~\ref{fig:gen}.
\begin{enumerate}
\item \emph{Drawing graphs between loop points.} All graphs connecting the $L$ loop points with at most one edge between two points are generated edge by edge, with a planarity test after every edge, and the nonplanar ones are discarded.
\item \emph{Attaching the external points.} Each graph connecting the loop points is joined in all possible ways to the $n$ external points, one edge at a time and again with at most one edge between two points. To enforce the cyclic order of the legs, the points $x_1,\dots,x_n$ are joined into a cycle and each to an auxiliary vertex. A graph is kept if it is planar and connected, every loop point carries at least four propagators, as dual conformal invariance requires, and every external point carries at least one.   Each kept graph defines the denominator of a term. An example of this procedure is represented in figure~\ref{fig:planarity}: graph (a) in the picture corresponds to the denominator
\begin{equation}
    \frac{1}{x^2_{A1}\,x^2_{A2}\,x^2_{A4}\,x^2_{AB}\,x^2_{B2}\,x^2_{B3}\,x^2_{B4}}
    \label{eq:den-a}
\end{equation}
and graph (b) to
\begin{equation}
    \frac{1}{x^2_{A1}\,x^2_{A2}\,x^2_{A3}\,x^2_{AB}\,x^2_{B2}\,x^2_{B3}\,x^2_{B4}}.
    \label{eq:den-b}
\end{equation}
While both may seem in principle acceptable, only the expression in eq.~\eqref{eq:den-a} represents a planar term: the denominator in eq.~\eqref{eq:den-b}, corresponding to a nonplanar graph, is discarded.

\begin{figure}[p]
\centering
\begin{tikzpicture}[>=Stealth, node distance=0.32cm,
  box/.style={draw,rectangle,align=center,font=\scriptsize,inner sep=3pt,text width=6.0cm},
  res/.style={draw,rectangle,rounded corners=6pt,fill=black!10,align=center,font=\scriptsize,inner sep=3pt,text width=5.4cm},
  dec/.style={draw,diamond,aspect=3.4,align=center,font=\scriptsize,inner sep=1pt},
  num/.style={draw,circle,fill=white,font=\scriptsize,inner sep=1pt,minimum size=0.36cm},
  lab/.style={font=\scriptsize},
  arr/.style={->,line width=0.5pt}]
\node[box] (g1) {for $n$ legs and $L$ loops:\\ generate all graphs on the $L$ loop points,\\ with at most one edge between two points};
\node[dec, below=of g1] (d1) {planar?};
\node[res, below=of d1] (o1) {graphs on the loop points};
\node[box, below=of o1] (g2) {join each graph in all possible ways to the\\ $n$ external points, in their cyclic order};
\node[dec, below=of g2] (d2) {planar and connected,\\ valence bounds met?};
\node[res, below=of d2] (o2) {dual graphs};
\node[box, below=of o2] (g3) {build each numerator that makes\\ the term dual conformally invariant};
\node[dec, below=of g3] (d3) {numerator\\ exists?};
\node[box, below=of d3] (g4) {compare the hash key of each new orbit\\ with those of the orbits already found};
\node[dec, below=of g4] (d4) {new orbit?};
\node[res, below=of d4] (o4) {terms (dual graph and numerator),\\ one for each orbit};
\node[dec, below=of o4] (d5) {sector?};
\node[box, below=of d5] (g5) {form all products of a term\\ with an $R$-invariant};
\node[box, below=of g5] (g6) {sum each term over all dihedral\\ and loop relabelings};
\node[box, below=of g6] (g7) {select a maximal linearly independent subset\\ by evaluation over a finite field};
\node[res, below=of g7] (o7) {ansatz};
\foreach \a/\b in {g1/d1,d1/o1,o1/g2,g2/d2,d2/o2,o2/g3,g3/d3,d3/g4,g4/d4,d4/o4,o4/d5,d5/g5,g5/g6,g6/g7,g7/o7} \draw[arr] (\a) -- (\b);
\foreach \d in {d1,d2,d3,d4} \node[lab,anchor=west] at ($(\d.south)+(0.08,-0.16)$) {yes};
\node[lab,anchor=west] at ($(d5.south)+(0.08,-0.16)$) {NMHV};
\foreach \d in {d1,d2,d3,d4} \draw[arr] (\d.east) -- ++(2.2,0) node[lab,above,pos=0.35] {no} node[lab,right] {discard};
\draw[arr] (d5.east) -- ++(4.2,0) node[lab,above,pos=0.5] {MHV} |- ($(g5.south)!0.5!(g6.north)$) -- (g6.north);
\foreach \n/\k in {g1/1,g2/2,g3/3,g4/4,g5/5,g6/6,g7/7} \node[num] at (\n.north west) {\k};
\path (-5.8,0) (5.8,0);
\end{tikzpicture}
\caption{Generation of the ansatz for the integrand with $n$ legs at $L$ loops. The numbered boxes are the steps of section~\ref{sec:ansatz}; in the MHV sector step 5 is skipped.}
\label{fig:gen}
\end{figure}
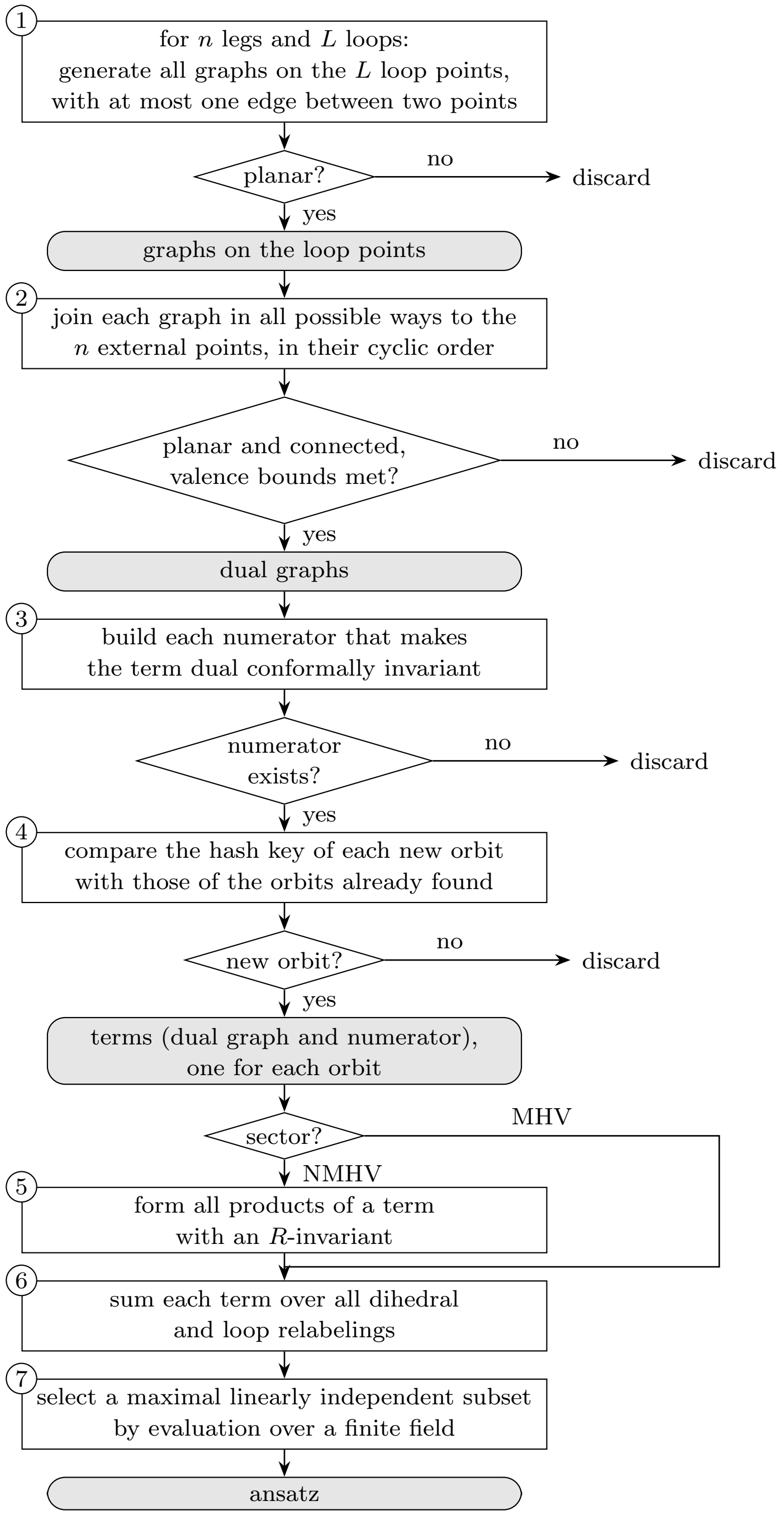

\item \emph{Building numerators.} For each graph, all numerators that are polynomials in four-brackets containing complete loop lines, and are compatible with dual conformal invariance, are generated. A graph that admits no numerator is discarded.
\item \emph{Erasing duplicates.} After a numerator is built, the program generates the dihedral orbit of the new term, sums the terms in it, and compares it with those of the terms already generated, discarding duplicates.
\item \emph{Multiplying by $R$-invariants.} In the NMHV sector all products of a term with an $R$-invariant of eq.~\eqref{eq:Rtw} are formed.
\item \emph{Symmetrizing.} Each term is replaced by the sum of its images under all elements of the group of dihedral and loop relabelings.
\item \emph{Selecting an independent subset.} The symmetrized terms are in general linearly dependent, through the Pl\"ucker relations among the four-brackets and, in the NMHV sector, through the six-term identity of the $R$-invariants. All terms are evaluated at random momentum twistors over a finite field, and a linearly independent subset of rational functions
is selected to form the ansatz.
\end{enumerate}
The outcome of this procedure is that the integrand is written
as a linear combination of a selected basis of rational functions,
with unknown rational coefficients.

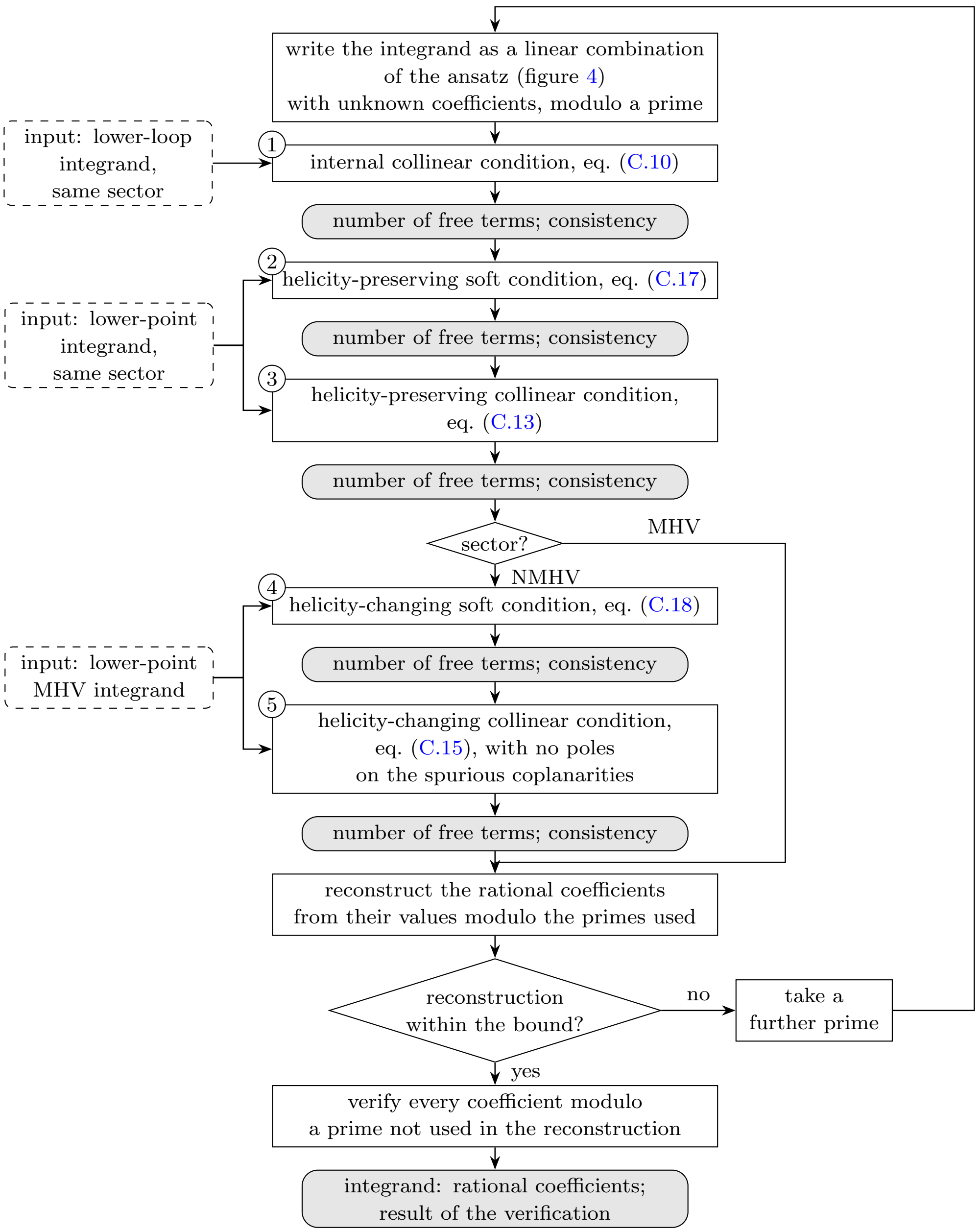
\begin{figure}[p]
\centering
\begin{tikzpicture}[>=Stealth, node distance=0.3cm,
  box/.style={draw,rectangle,align=center,font=\scriptsize,inner sep=3pt,text width=5.8cm},
  res/.style={draw,rectangle,rounded corners=6pt,fill=black!10,align=center,font=\scriptsize,inner sep=3pt,text width=5.0cm},
  dec/.style={draw,diamond,aspect=3.2,align=center,font=\scriptsize,inner sep=1pt},
  inp/.style={draw,dashed,rectangle,rounded corners=3pt,align=center,font=\scriptsize,inner sep=3pt,text width=2.6cm},
  num/.style={draw,circle,fill=white,font=\scriptsize,inner sep=1pt,minimum size=0.36cm},
  lab/.style={font=\scriptsize},
  arr/.style={->,line width=0.5pt}]
\node[box] (s0) {write the integrand as a linear combination\\ of the ansatz (figure~\ref{fig:gen})\\ with unknown coefficients, modulo a prime};
\node[box, below=of s0] (c1) {internal collinear condition, eq.~\eqref{eq:twcond}};
\node[res, below=of c1] (r1) {number of free terms; consistency};
\node[box, below=of r1] (c2) {helicity-preserving soft condition, eq.~\eqref{eq:hp-soft-cond}};
\node[res, below=of c2] (r2) {number of free terms; consistency};
\node[box, below=of r2] (c3) {helicity-preserving collinear condition,\\ eq.~\eqref{eq:hp-coll-cond}};
\node[res, below=of c3] (r3) {number of free terms; consistency};
\node[dec, below=of r3] (ds) {sector?};
\node[box, below=of ds] (c4) {helicity-changing soft condition, eq.~\eqref{eq:hc-soft-cond}};
\node[res, below=of c4] (r4) {number of free terms; consistency};
\node[box, below=of r4] (c5) {helicity-changing collinear condition,\\ eq.~\eqref{eq:hc-coll-cond}, with no poles\\ on the spurious coplanarities};
\node[res, below=of c5] (r5) {number of free terms; consistency};
\node[box, below=of r5] (rc) {reconstruct the rational coefficients\\ from their values modulo the primes used};
\node[dec, below=of rc] (dr) {reconstruction\\ within the bound?};
\node[box, text width=1.9cm, right=1.0cm of dr] (np) {take a\\ further prime};
\node[box, below=of dr] (vf) {verify every coefficient modulo\\ a prime not used in the reconstruction};
\node[res, below=of vf] (fin) {integrand: rational coefficients;\\ result of the verification};
\foreach \a/\b in {s0/c1,c1/r1,r1/c2,c2/r2,r2/c3,c3/r3,r3/ds,ds/c4,c4/r4,r4/c5,c5/r5,r5/rc,rc/dr,dr/vf,vf/fin} \draw[arr] (\a) -- (\b);
\node[lab,anchor=west] at ($(ds.south)+(0.08,-0.16)$) {NMHV};
\node[lab,anchor=west] at ($(dr.south)+(0.08,-0.16)$) {yes};
\draw[arr] (ds.east) -- ++(3.0,0) node[lab,above,pos=0.5] {MHV} |- ($(r5.south)!0.5!(rc.north)$) -- (rc.north);
\draw[arr] (dr.east) -- (np.west) node[lab,above,pos=0.5] {no};
\draw[arr] (np.east) -- ++(1.1,0) |- ($(s0.north)+(0,0.35)$) -- (s0.north);
\node[inp, left=0.8cm of c1] (i1) {input: lower-loop\\ integrand, same sector};
\draw[arr] (i1.east) -- (c1.west);
\node[inp] (i2) at ($(c2.west)!0.5!(c3.west)+(-2.2,0)$) {input: lower-point\\ integrand, same sector};
\draw[arr] (i2.east) -- ++(0.4,0) |- (c2.west); \draw[arr] (i2.east) -- ++(0.4,0) |- (c3.west);
\node[inp] (i3) at ($(c4.west)!0.5!(c5.west)+(-2.2,0)$) {input: lower-point\\ MHV integrand};
\draw[arr] (i3.east) -- ++(0.4,0) |- (c4.west); \draw[arr] (i3.east) -- ++(0.4,0) |- (c5.west);
\foreach \n/\k in {c1/1,c2/2,c3/3,c4/4,c5/5} \node[num] at (\n.north west) {\k};
\end{tikzpicture}
\caption{Application of the conditions to the ansatz for the integrand with $n$ legs at $L$ loops. The numbered boxes are the conditions. In the MHV sector conditions 4 and 5 are skipped.}
\label{fig:flow}
\end{figure}
\subsection{Application of the conditions}
\label{sec:application}
The unknown coefficients that appear in the ansatz are determined by imposing a set of conditions that characterize specific limits in the space of external kinematics and loop momenta, as represented in figure~\ref{fig:flow}.
\begin{itemize}
    \item The \textit{internal collinear condition}, {eq.}~\eqref{eq:intro-condition}. This relates the integrand at $L$ loops, in the limit where a loop momentum becomes collinear with an external one, to the one at $(L-1)$ loops.
    \item The \textit{helicity-preserving soft condition}~\eqref{eq:hp-soft-dual} relates an integrand of multiplicity $n$, in the limit where the momentum of one leg becomes soft, to the one of multiplicity $(n-1)$ in the same helicity sector. The \textit{helicity-changing soft condition}~\eqref{eq:hc-soft-dual}, expressed in momentum twistors in {eq.}~\eqref{eq:hc-soft-cond}, relates an N$^k$MHV integrand of multiplicity $n$ to the one of multiplicity $(n-1)$ in the N$^{k-1}$MHV sector.
    \item The \textit{helicity-preserving collinear condition}~\eqref{eq:hp-coll-dual} and \textit{helicity-changing collinear condition}~\eqref{eq:hc-coll-dual} do the same as the respective soft conditions, but for the limit in which the momenta of two external legs become collinear with each other. These include the vanishing of nonphysical poles that appear in the $R$-invariants.
    \item The \textit{subleading soft condition} is the statement that the subleading order in the helicity-preserving soft limit~\eqref{eq:hp-soft-dual} must vanish. We impose it separately as a check on the integrands.
\end{itemize}

Each condition constrains the integrand on a set of kinematic configurations, the locus of the condition: the collinear pole of a loop line for the internal collinear condition, and the parameterized soft or collinear limit for the external conditions. The conditions are imposed at random points of their loci. In the NMHV sector the Grassmann variables $\eta_i^A$ are also replaced by random commuting numbers. Each condition gives a set of linear equations for the coefficients. Its right-hand side is computed from the lower-loop and lower-point integrands, or vanishes. All linear systems are solved over finite fields, modulo primes close to $2^{31}$, and the rational values of coefficients are then reconstructed.

\section{Results}
\label{sec:results}

We applied the procedure of section~\ref{sec:bootstrap} to the MHV and NMHV integrands with $n+L\le9$, MHV integrands with $n>4$ and $n+L=10$, and the five-point five-loop, the six-point four-loop and the nine-point one-loop NMHV integrands.

\subsection{Implementation and performance}
\label{sec:implementation}
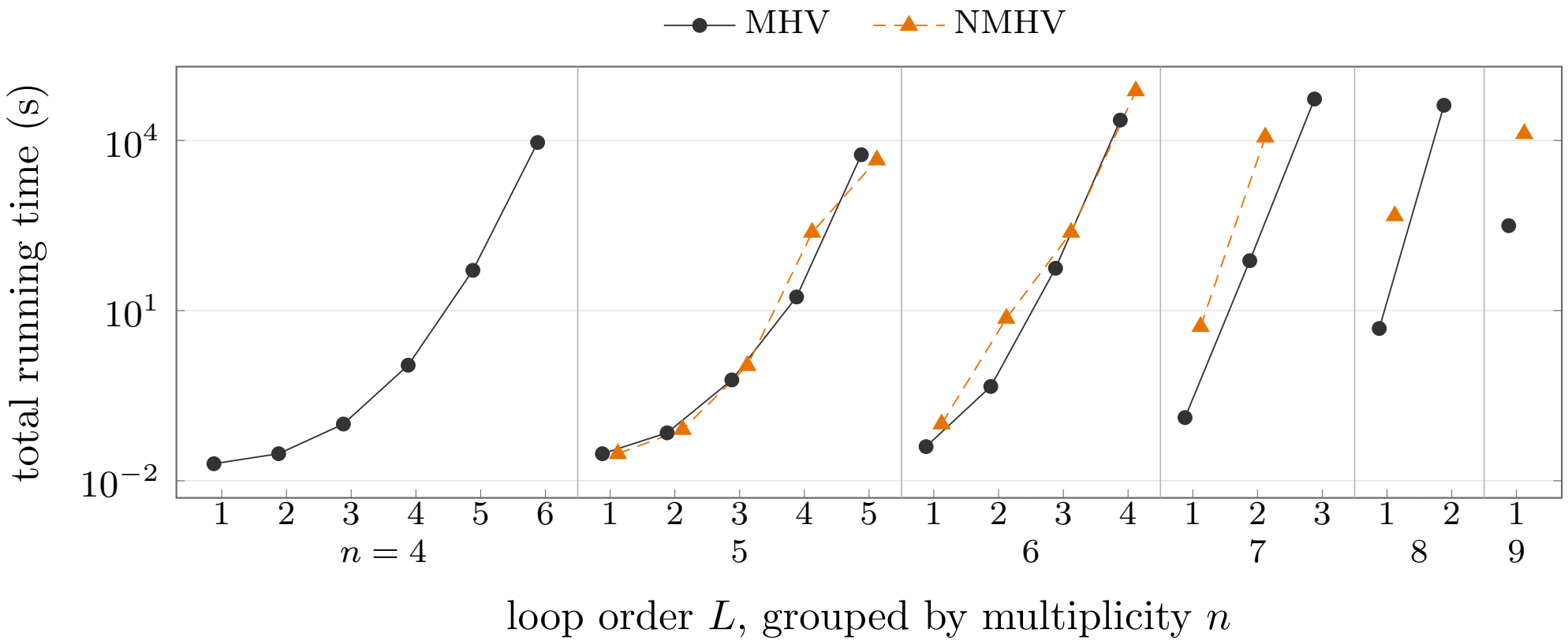
\begin{figure}[t]
\centering
\pgfplotsset{
  bench/.style={width=13.2cm, height=5.2cm, xmin=0.3, xmax=21.7,
    ymajorgrids=true, grid style={line width=0.2pt, draw=black!14}, axis line style={line width=0.4pt, draw=black!60},
    tick style={draw=black!60}, tick label style={font=\footnotesize}, label style={font=\small},
    xtick={1,...,21}, xticklabels={1,2,3,4,5,6,1,2,3,4,5,1,2,3,4,1,2,3,1,2,1}, x tick label style={font=\scriptsize, inner sep=1pt}, xtick pos=bottom, major tick length=2pt,
    extra x ticks={3.5,9,13.5,17,19.5,21}, extra x tick labels={$n=4$,$5$,$6$,$7$,$8$,$9$}, extra x tick style={major tick length=0pt, tick label style={font=\footnotesize, yshift=-1.9ex}},
    xlabel={loop order $L$, grouped by multiplicity $n$}, xlabel style={yshift=-0.9ex}, unbounded coords=jump, legend cell align=left,
    legend style={font=\scriptsize, draw=none, fill=white, row sep=-1pt, inner sep=1pt}},
  tm/.style={mark=*, mark size=1.6pt, color=black!80, line width=0.35pt, mark options={fill=black!80}},
  tn/.style={mark=triangle*, mark size=2.2pt, color=orange!90!black, line width=0.35pt, densely dashed, mark options={solid, fill=orange!90!black}}}
\begin{tikzpicture}
\begin{axis}[bench, ymode=log, ylabel={total running time (s)}, ymin=0.005, ymax=200000, name=ax1, legend columns=2, legend style={at={(0.5,1.03)}, anchor=south, /tikz/every even column/.append style={column sep=8pt}}]
\addplot[tm] coordinates {(0.88,0.02) (1.88,0.03) (2.88,0.1) (3.88,1.09) (4.88,51.34) (5.88,9208) (6.5,nan) (6.88,0.03) (7.88,0.07) (8.88,0.6) (9.88,17.42) (10.88,5597) (11.5,nan) (11.88,0.04) (12.88,0.46) (13.88,55.82) (14.88,2.281e+04) (15.5,nan) (15.88,0.13) (16.88,75.86) (17.88,5.379e+04) (18.5,nan) (18.88,4.84) (19.88,4.172e+04) (20.5,nan) (20.88,314.1)}; \addlegendentry{MHV}
\addplot[tn] coordinates {(7.12,0.03) (8.12,0.08) (9.12,1.07) (10.12,239.1) (11.12,4517) (11.5,nan) (12.12,0.1) (13.12,7.29) (14.12,239.1) (15.12,7.46e+04) (15.5,nan) (16.12,5.23) (17.12,1.145e+04) (18.5,nan) (19.12,467.7) (20.5,nan) (21.12,1.313e+04)}; \addlegendentry{NMHV}
\draw[black!30, line width=0.3pt] (axis cs:6.5,0.005) -- (axis cs:6.5,200000);
\draw[black!30, line width=0.3pt] (axis cs:11.5,0.005) -- (axis cs:11.5,200000);
\draw[black!30, line width=0.3pt] (axis cs:15.5,0.005) -- (axis cs:15.5,200000);
\draw[black!30, line width=0.3pt] (axis cs:18.5,0.005) -- (axis cs:18.5,200000);
\draw[black!30, line width=0.3pt] (axis cs:20.5,0.005) -- (axis cs:20.5,200000);
\end{axis}
\end{tikzpicture}
\caption{Total running time of the computation of each integrand $(n,L)$, on 30 cores of a node with two Intel Xeon Silver 4216 processors at 2.1\,GHz and 187\,GB. The running time is the sum of the generation of the ansatz and of the application of the conditions. }
\label{fig:benchmark}
\end{figure}
A first implementation of the procedure of section~\ref{sec:bootstrap} in Mathematica determined the integrands up to $n+L=8$. An implementation in C++, written with the help of the AI model Claude by Anthropic, reached the remaining integrands with $n+L\le10$; all results of this section were obtained with it. {The code is available from the authors upon request.}
 Figure~\ref{fig:benchmark} shows, for every integrand and separately for the MHV and NMHV sectors, the total running time. Peak memory usage followed the same trend, and ranged from $10^{-2}$ to $10^2$ GB in the cases considered.

The six-point three-loop MHV integrand illustrates the sizes involved. Steps 1 and 2 of section~\ref{sec:ansatz} produce 1320 dual graphs, which form 27 different terms up to dihedral and loop relabelings; steps 3 and 4 produce 7837 orbits of numerators under these relabelings; finally, step 7 selects {674} independent elements. Of these, the internal collinear condition leaves 24 free, the helicity-preserving soft condition 10, and the helicity-preserving collinear condition none. The resulting integrand has 287 nonzero coefficients, whose numerators are at most 31 in absolute value and whose denominators are at most 6. The generation of the ansatz takes about 52 seconds on 30 cores of the machine of figure~\ref{fig:benchmark}, and the application of the conditions takes an additional 4 seconds.

\subsection{Determination of the integrands and strength of each condition}
\label{sec:strength}
\begin{table}[htbp]
\centering
\small\setlength{\tabcolsep}{3.4pt}
\begin{tabular}{cl|cccccc}
\toprule
$n$ & & \multicolumn{6}{c}{loop order $L$}\\
& & 1 & 2 & 3 & 4 & 5 & 6\\
\midrule
4 & size of the ansatz & 1 & 1 & 2 & 10 & 74 & 1142\\
 & internal collinear & 0\,(0) & 0\,(0) & 0\,(0) & 2\,(2) & 18\,(18) & 351\,(351)\\
 & helicity-preserving soft & 0\,(1) & 0\,(1) & 0\,(2) & 2\,(10) & 18\,(74) & 323\,(1110)\\
 & helicity-preserving collinear & 0\,(1) & 0\,(1) & 0\,(2) & 0\,(8) & 3\,(49) & 10\,(506)\\
 & subleading soft & --\,(1) & --\,(1) & --\,(2) & --\,(10) & --\,(74) & --\,(984)\\
\midrule
5 & size of the ansatz & 2 & 4 & 22 & 219 & 3545 & \\
 & internal collinear & 1\,(1) & 0\,(0) & 0\,(0) & 19\,(19) & 471\,(471) & \\
 & helicity-preserving soft & 1\,(1) & 0\,(2) & 0\,(14) & 12\,(132) & 211\,(1981) & \\
 & helicity-preserving collinear & 0\,(0) & 0\,(0) & 0\,(0) & 0\,(0) & 0\,(5) & \\
 & subleading soft & --\,(1) & --\,(1) & --\,(2) & --\,(5) & --\,(43) & \\
\midrule
6 & size of the ansatz & 9 & 66 & 674 & 9895 & & \\
 & internal collinear & 3\,(3) & 5\,(5) & 24\,(24) & 535\,(535) & & \\
 & helicity-preserving soft & 0\,(0) & 2\,(11) & 10\,(147) & 95\,(2189) & & \\
 & helicity-preserving collinear & 0\,(0) & 0\,(0) & 0\,(0) & 0\,(0) & & \\
 & subleading soft & --\,(1) & --\,(1) & --\,(2) & --\,(5) & & \\
\midrule
7 & size of the ansatz & 29 & 871 & 18019 & & & \\
 & internal collinear & 13\,(13) & 60\,(60) & 574\,(574) & & & \\
 & helicity-preserving soft & 0\,(0) & 1\,(9) & 17\,(626) & & & \\
 & helicity-preserving collinear & 0\,(0) & 0\,(0) & 0\,(0) & & & \\
 & subleading soft & --\,(1) & --\,(1) & --\,(2) & & & \\
\midrule
8 & size of the ansatz & 127 & 11988 & & & & \\
 & internal collinear & 65\,(65) & 929\,(929) & & & & \\
 & helicity-preserving soft & 0\,(0) & 1\,(17) & & & & \\
 & helicity-preserving collinear & 0\,(0) & 0\,(0) & & & & \\
 & subleading soft & --\,(1) & --\,(1) & & & & \\
\midrule
9 & size of the ansatz & 463 & & & & & \\
 & internal collinear & 263\,(263) & & & & & \\
 & helicity-preserving soft & 0\,(0) & & & & & \\
 & helicity-preserving collinear & 0\,(0) & & & & & \\
 & subleading soft & --\,(1) & & & & & \\
\bottomrule
\end{tabular}
\caption{MHV sector: number of terms of the ansatz left free by the conditions. For each multiplicity $n$, the first row gives the size of the ansatz at each loop order $L$. Each further row gives the number of terms left free by the condition named in it together with those of the rows above and, in parentheses, by that condition alone. The last row, corresponding to the subleading soft condition, only records the count where it is imposed alone.}
\label{tab:mhv}
\end{table}

\begin{table}[htbp]
\centering
\small\setlength{\tabcolsep}{3.4pt}
\begin{tabular}{cl|ccccc}
\toprule
$n$ & & \multicolumn{5}{c}{loop order $L$}\\
& & 1 & 2 & 3 & 4 & 5\\
\midrule
5 & size of the ansatz & 2 & 4 & 22 & 219 & 3545\\
 & internal collinear & 1\,(1) & 0\,(0) & 0\,(0) & 19\,(19) & 471\,(471)\\
 & helicity-preserving soft & 1\,(2) & 0\,(4) & 0\,(22) & 19\,(219) & 471\,(3545)\\
 & helicity-preserving collinear & 1\,(2) & 0\,(4) & 0\,(19) & 10\,(156) & 171\,(2080)\\
 & helicity-changing soft & 1\,(1) & 0\,(2) & 0\,(14) & 5\,(132) & 101\,(1981)\\
 & helicity-changing collinear & 0\,(0) & 0\,(0) & 0\,(0) & 0\,(0) & 0\,(6)\\
 & subleading soft & --\,(2) & --\,(4) & --\,(22) & --\,(219) & --\,(3545)\\
\midrule
6 & size of the ansatz & 26 & 268 & 3110 & 48407 & \\
 & internal collinear & 7\,(7) & 13\,(13) & 81\,(81) & 2506\,(2506) & \\
 & helicity-preserving soft & 3\,(10) & 9\,(143) & 51\,(1902) & 1418\,(31140) & \\
 & helicity-preserving collinear & 1\,(4) & 1\,(39) & 4\,(425) & 76\,(5958) & \\
 & helicity-changing soft & 1\,(7) & 1\,(120) & 4\,(1637) & 45\,(26912) & \\
 & helicity-changing collinear & 0\,(0) & 0\,(1) & 0\,(24) & 0\,(557) & \\
 & subleading soft & --\,(2) & --\,(3) & --\,(90) & --\,(3001) & \\
\midrule
7 & size of the ansatz & 339 & 12621 & & & \\
 & internal collinear & 129\,(129) & 798\,(798) & & & \\
 & helicity-preserving soft & 2\,(4) & 88\,(1363) & & & \\
 & helicity-preserving collinear & 0\,(0) & 0\,(57) & & & \\
 & helicity-changing soft & 0\,(133) & 0\,(6606) & & & \\
 & helicity-changing collinear & 0\,(32) & 0\,(1488) & & & \\
 & subleading soft & --\,(2) & --\,(3) & & & \\
\midrule
8 & size of the ansatz & 3529 & & & & \\
 & internal collinear & 1671\,(1671) & & & & \\
 & helicity-preserving soft & 0\,(0) & & & & \\
 & helicity-preserving collinear & 0\,(0) & & & & \\
 & helicity-changing soft & 0\,(1846) & & & & \\
 & helicity-changing collinear & 0\,(1162) & & & & \\
 & subleading soft & --\,(2) & & & & \\
\midrule
9 & size of the ansatz & 29763 & & & & \\
 & internal collinear & 16467\,(16467) & & & & \\
 & helicity-preserving soft & 0\,(0) & & & & \\
 & helicity-preserving collinear & 0\,(0) & & & & \\
 & helicity-changing soft & 0\,(18505) & & & & \\
 & helicity-changing collinear & 0\,(14662) & & & & \\
 & subleading soft & --\,(2) & & & & \\
\bottomrule
\end{tabular}
\caption{As table~\ref{tab:mhv}, for the NMHV sector.}
\label{tab:nmhv}
\end{table}
Tables~\ref{tab:mhv} and~\ref{tab:nmhv} show how the conditions reduce the ansatz in the MHV and NMHV sectors. Recall that we determine integrands by applying conditions recursively in $(n,L)$: in all counts the lower-loop and lower-point integrands on the right-hand sides of the conditions are set to their true values. Each entry of the table is therefore the dimension of the kernel of the listed conditions on the ansatz with $n$ legs at $L$ loops, so that no freedom propagates from lower $n$ or $L$.

Let us discuss in detail the MHV integrands at four points. Through four loops, the conditions fix the four-point integrand; at five and six loops they do not. At $(4,5)$ the conditions leave three terms undetermined. Kept as unknowns in the five-point conditions at the same loop order, whose right-hand sides contain the four-point integrand, two of them are fixed by the five-point MHV conditions and the third by the helicity-changing collinear condition on the five-point NMHV integrand. At $(4,6)$ the internal collinear condition leaves 351 terms, and the helicity-preserving soft and collinear conditions determine all of them except 10. {In \cite{Bourjaily:2011hi}, these terms were fixed by the additional condition that the integrand should be finite when external legs are taken to be off-shell, which we do not apply. Nevertheless,} we expect the six-loop five-point integrands, beyond our range, to fix these remaining degrees of freedom.

In general, relative to the size of the ansatz, the internal collinear condition gains strength with the loop order and loses it with the multiplicity, as observed in ref.~\cite{Golden:2012hi}; without it, and applying only the limits in external kinematics, multiple integrands remain undetermined both in the MHV and NMHV sectors. The helicity-changing conditions lose strength with both loop order and multiplicity, and helicity-preserving conditions instead gain strength as the multiplicity grows. After the application of all conditions in sequence, with the exception of the four-point, six-loop case, all terms are uniquely fixed.

\subsection{Checks on the procedure and on the solutions}
\label{sec:checks}

We performed checks on the procedure and on the integrands it determines.
\begin{itemize}
\item \emph{Internal consistency of the procedure.} Every linear system we solve is overdetermined, and each was verified to have a solution.
\item \emph{Subleading soft condition.} Every integrand satisfies the subleading soft condition, i.e.\ the subleading order of its soft limit vanishes.
\item \emph{Consistency with parity.} The five-point NMHV integrands coincide with the parity images of the five-point MHV ones.
\item \emph{Agreement with BCFW recursion.} All our results for the integrands agree with those obtained with the BCFW-based loop-integrand recursion of ref.~\cite{Bourjaily:2023apy}, computed using the Mathematica package provided there. Both the recursion and our integrands were evaluated at three random rational points, and they agree after taking into account a conventional sign difference {of $(-1)^L$, coming from the sign of the normalization in our definition \eqref{eq:defint}}. The integrands were also compared with the local forms available in the literature, up to three loops in the MHV sector and to two loops for NMHV~\cite{Arkani-Hamed:2010pyv}.
\item \emph{Agreement with known local representations.} Up to five loops, the five-point integrand was checked to coincide with the one in ref.~\cite{Ambrosio:2013pba}, and the four-point integrand with that of ref.~\cite{Bourjaily:2011hi}. We verified this by evaluation over finite fields at three random kinematic points. {We also checked that the known four-point, six-loop integrand {satisfies} all the conditions we impose.}
\end{itemize}
Every integrand we obtained passed all the checks that apply to it.

\subsection{New results for integrands in local form}
\label{sec:local}

We provide all the integrands we determined in machine-readable form. They are given in the ancillary files, one file for each sector and each pair $(n,L)$, as linear combinations of the elements of the ansatz with rational coefficients, expressed in momentum twistor variables. While the integrand is in principle obtainable at all loops by recursion, in terms of a sum whose individual components have nonphysical poles, the local forms we obtain are built from bases that are free of these spurious divergences, allowing in principle for term-by-term regularization and integration. We obtain new results in local form for the six-point three-loop NMHV integrand, and for the six-point four-loop MHV and NMHV integrands, for which, to the best of our knowledge, no earlier representation of this kind appears in the literature.

\section{Discussion and outlook}
\label{sec:outlook}

In this paper we derived the condition~\eqref{eq:intro-condition} that the loop integrand of planar $\mathcal{N}=4$ sYM satisfies when a loop momentum becomes collinear with an external leg. The derivation holds in every helicity sector. {A relation of this form is expected from the all-order analysis of collinear factorization of ref.~\cite{Feige:2014wja}, as discussed in the introduction. For MHV integrands it is equivalent to the observation, used in refs.~\cite{Bourjaily:2011hi,Golden:2012hi}, that} the integrand of the logarithm of the amplitude has no double pole when a single loop momentum becomes collinear with an external leg.

Together with the soft and collinear limits of the external kinematics, and other assumptions listed in section~\ref{sec:ansatz}, this internal collinear condition fixes the MHV and NMHV integrands uniquely for all the $n$ and $L$ with $n+L\le9$, as well as all MHV integrands with $n>4$ and $n+L=10$, and the five-point five-loop, the six-point four-loop and the nine-point one-loop NMHV integrands. This includes new local forms of the six-point four-loop MHV and NMHV integrands, and the six-point three-loop NMHV integrand. All the integrands we obtained agree with those available in the literature.

Our results show that the infrared behavior of the planar loop integrand, a consequence of its factorization, carries enough information to fix it over a wide range of multiplicities and loop orders.

We see several directions for further work.
\begin{enumerate}
\item \textit{N$^2$MHV and higher sectors.} All the conditions we use apply unchanged in these sectors, with the helicity-changing limits relating the N$^k$MHV integrand to the N$^{k-1}$MHV one. The extension requires a basis of the $R$-invariants that multiply the integrand, which are known: they appear in the solution of the tree-level recursion~\cite{Drummond:2008cr} and take a simple form in momentum twistors~\cite{Mason:2009qx}.
\item \textit{Extension to nonplanar amplitudes.} The method described in this paper relies on the planar loop integrand. Beyond the planar limit, a loop integrand must first be defined. Definitions of canonical integrands through periodic dual variables~\cite{Ben-Israel:2018ckc}, the formulation of amplitudes through curves on surfaces~\cite{Arkani-Hamed:2023lbd,Arkani-Hamed:2023jry}, as well as global loop-integrand bases such as those constructed in ref.~\cite{Bern:2024vqs}, may provide the canonical object on which conditions of this kind could be imposed. Form factors of the operator $\mathrm{tr}\,\phi^2$, the lowest component of the stress-tensor multiplet, are another interesting case: even in the 't Hooft limit they contain nonplanar integrals, and much is known about them (see ref.~\cite{Yang:2019hkn} for a review).
\item \textit{Pure Yang--Mills theory.} The internal collinear condition follows from the factorization of the integrand on its poles, so that analogous conditions exist in any theory whose loop integrand factorizes, and in particular for appropriately defined integrands of pure Yang--Mills theory. There, some of the assumptions that underpin our construction of integrands in $\mathcal{N}=4$ sYM may no longer hold and would need to be revised accordingly. In that context, it would be especially interesting to extend the procedure described here from four-dimensional to $D$-dimensional integrands. {For a class of QCD amplitudes, integrands whose infrared singularities factorize locally have been constructed through two loops~\cite{Anastasiou:2020sdt,Anastasiou:2022eym}.}
\end{enumerate}

\acknowledgments
{The authors thank Jacob Bourjaily and Simon Caron-Huot for useful correspondence and feedback.} JMH thanks Simon Caron-Huot for initial joint work on the factorization argument of section~\ref{sec:condition}. DC\ thanks Paola Cappellazzi, S\'ergio Carr\^olo, Lance J.~Dixon, Jin Dong, Ben Hollering, Marcus B.~Spradlin, Jaroslav Trnka, Cristian Vergu and Anastasia Volovich for stimulating discussions. The AI model Claude by
Anthropic was used for code implementation and in the editing process of this manuscript. The authors take full responsibility for all contents of this paper. This project was supported by the European Research Council (ERC) under the European Union's Horizon Europe research and innovation programme (grant agreement No.~101118787, UNIVERSE+), and by the Deutsche Forschungsgemeinschaft (DFG, German Research Foundation) through the research unit FOR 5582 -- Projektnummer 508889767. Views and opinions expressed are however those of the authors only and do not necessarily reflect those of the European Union or the European Research Council Executive Agency. Neither the European Union nor the granting authority can be held responsible for them.

\appendix
\section{Tree-level splitting amplitudes}
\label{app:split}

As two adjacent legs $a$ and $b$ of a tree amplitude become collinear, $a\to zP$ and $b\to(1-z)P$ with $P=a+b$ and $z$ the momentum fraction, the amplitude behaves universally~\cite{Mangano:1990by},
\begin{equation}
\mathcal{A}_{n+1}^{\text{tree}}\big[\dots,a^{I_1},b^{I_2},\dots\big]\;\longrightarrow\;\sum_{I_3}\mathrm{Split}^{\text{tree}}_{\bar I_3}\big[z,\,a^{I_1},b^{I_2}\big]\;\mathcal{A}_n^{\text{tree}}\big[\dots,P^{I_3},\dots\big]\,,
\label{eq:splitdef}
\end{equation}
which defines the splitting amplitude. Comparing with the factorization on the pole $s_{ab}=P^2\to0$ gives it as a three-point amplitude in the collinear kinematics divided by the invariant of the channel,
\begin{equation}
\mathrm{Split}^{\text{tree}}_{\bar I_3}\big[z,\,a^{I_1},b^{I_2}\big]\;=\;\lim_{s_{ab}\to0}\frac{\mathcal{A}_3^{\text{tree}}\big[a^{I_1},b^{I_2},-P^{\bar I_3}\big]}{s_{ab}}\,.
\label{eq:splitA3}
\end{equation}
The spinor conventions are $2\,a\cdot b=\langle ab\rangle[ba]$, $|v\rangle=-|{-v}\rangle$ and $|v]=|{-v}]$. For complex momenta the three-point amplitude in eq.~\eqref{eq:splitA3}, and with it the splitting amplitude, is MHV when the square spinors of $a$ and $b$ are aligned and $\MHVb$ when their angle spinors are.

In superspace the sum over the states $I_3$ of the exchanged line becomes an integral over its Grassmann variables, and the three-point superamplitudes are
\begin{equation}
\begin{aligned}
\mathcal{A}_3^{\text{MHV}}[1,2,3;\eta_1,\eta_2,\eta_3]&=\frac{1}{\langle12\rangle\langle23\rangle\langle31\rangle}\,\delta^{0|8}\big(\lambda_1 \eta_1+\lambda_2 \eta_2+\lambda_3 \eta_3\big)\,,\\
\mathcal{A}_3^{\MHVb}[1,2,3;\eta_1,\eta_2,\eta_3]&=\frac{1}{[21][32][13]}\,\delta^{0|4}\big([21]\eta_3+[32]\eta_1+[13]\eta_2\big)\,.
\end{aligned}
\label{eq:A3super}
\end{equation}
The delta function of the MHV amplitude imposes the conservation of the supermomentum $\sum_i\lambda_i\eta_i$.

\section{Conditions on external kinematics}
\label{sec:external}

Integrand factorization also fixes the integrand as two adjacent external momenta become collinear or one becomes soft: the tree-normalized integrand reduces to the one with one leg fewer, at the same loop order~\cite{Drummond:2010mb}. Each limit has a helicity-preserving channel and a helicity-changing one, which lowers the helicity sector by one unit.
\begin{itemize}
\item \emph{Helicity-preserving collinear condition.} The angle spinors of $p_{n-1}$ and $p_n$ align, that is, $\langle n{-}1\,n\rangle\propto\delta_c$ vanishes with $[n{-}1\,n]$ finite, the channel of the three-point $\MHVb$ amplitude. Then
\begin{equation}
I_{n,k}^{(L)}\big(x_1,\dots,x_n;\,\theta_1,\dots,\theta_n\big)=I_{n-1,k}^{(L)}\big(x_1,\dots,x_{n-1};\,\theta_1,\dots,\theta_{n-1}\big)+\mathcal{O}(\delta_c)\,.
\label{eq:hp-coll-dual}
\end{equation}
\item \emph{Helicity-changing collinear condition.} The square spinors align instead, $[n{-}1\,n]\propto\delta_c\to0$ with $\langle n{-}1\,n\rangle$ finite, the channel of the three-point MHV amplitude. This amplitude has Grassmann degree eight, so that the integrand factorizes onto the $(n{-}1)$-point integrand one sector lower, times the tree-level ratio function, and for $k=1$
\begin{equation}
\begin{aligned}
I_{n,1}^{(L)}\big(x_1,\dots,x_n;\,\theta_1,\dots,\theta_n\big)={}&\mathcal{R}_{n,1}^{\text{tree}}\big(x_1,\dots,x_n;\,\theta_1,\dots,\theta_n\big)\Big|_{\delta_c^{-1}}\\
&\times I_{n-1,0}^{(L)}\big(x_1,\dots,x_{n-1}\big)+\mathcal{O}(\delta_c^0)\,,
\end{aligned}
\label{eq:hc-coll-dual}
\end{equation}
where $\mathcal{R}_{n,1}^{\text{tree}}=\mathcal{A}_{n,1}^{\text{tree}}/\mathcal{A}_{n,0}^{\text{tree}}$ is the NMHV ratio function, a sum of $R$-invariants~\eqref{eq:Rinv}, and $|_{\delta_c^{-1}}$ denotes its pole term, from the factor $x_{st}^2=x_{n-1,1}^2\propto\delta_c$ of the invariants that contain the channel. Analogous conditions relate the N$^k$MHV and N$^{k-1}$MHV sectors for every $k\ge1$; only $k=1$ is used here. The condition also requires the integrand to be regular on the spurious poles of the $R$-invariants, the four factors $\langle r|x_{rs}x_{st}|t\rangle$, $\langle r|x_{rs}x_{st}|t{-}1\rangle$, $\langle r|x_{rt}x_{ts}|s\rangle$ and $\langle r|x_{rt}x_{ts}|s{-}1\rangle$ of eq.~\eqref{eq:Rinv}.
\item \emph{Helicity-preserving soft condition.} The momentum $p_n\propto\delta_s$ vanishes with its Grassmann variable, so that $x_n\to x_1$ and $\theta_n\to\theta_1$, and
\begin{equation}
I_{n,k}^{(L)}\big(x_1,\dots,x_n;\,\theta_1,\dots,\theta_n\big)=I_{n-1,k}^{(L)}\big(x_1,\dots,x_{n-1};\,\theta_1,\dots,\theta_{n-1}\big)+\mathcal{O}(\delta_s^2)\,.
\label{eq:hp-soft-dual}
\end{equation}
 The vanishing of the term of order $\delta_s$ is the subleading soft theorem~\cite{Bern:2014oka,Cachazo:2014dia,Bianchi:2014gla}.
\item \emph{Helicity-changing soft condition.} The momentum $p_n\propto\delta_s$ is sent to zero, with the Grassmann variable $\eta_n$ of its leg held fixed. The $R$-invariants that contain the legs $n{-}1$, $n$ and $1$ then have a double pole. This is the soft limit of a negative-helicity gluon~\cite{Drummond:2010mb}, and for $k=1$
\begin{equation}
\begin{aligned}
I_{n,1}^{(L)}\big(x_1,\dots,x_n;\,\theta_1,\dots,\theta_n\big)={}&\mathcal{R}_{n,1}^{\text{tree}}\big(x_1,\dots,x_n;\,\theta_1,\dots,\theta_n\big)\Big|_{\delta_s^{-2}}\\
&\times I_{n-1,0}^{(L)}\big(x_1,\dots,x_{n-1}\big)+\mathcal{O}(\delta_s^{-1})\,.
\end{aligned}
\label{eq:hc-soft-dual}
\end{equation}
\end{itemize}

\section{Momentum twistors and the twistor form of the conditions}
\label{app:twistorform}

The ansatz of section~\ref{sec:bootstrap} is written, and the conditions of section~{\ref{sec:application}} are imposed, in momentum twistor variables. This appendix recalls these variables and gives the form of the integrand and of the conditions in them.

\subsection{The integrand in momentum twistor variables}
\label{sec:twistors}

 For the purpose of writing down a dual conformal ansatz for integrands, it is convenient to pass to momentum twistor variables, elements of a three-dimensional projective space which linearize the dual conformal transformations. Points of dual space correspond to lines in twistor space, each specified by a pair of twistors. Integrands are written~\cite{Arkani-Hamed:2010pyv,Arkani-Hamed:2010zjl} through the four-brackets
\begin{equation}
\langle ijkl\rangle \;=\; \epsilon_{ABCD}\,Z_i^A Z_j^B Z_k^C Z_l^D\,,
\label{eq:fourbracket}
\end{equation}
where the $Z$'s are points in a four-dimensional embedding of the projective space. This representation has the additional advantage of automatically enforcing momentum conservation as well as on-shell conditions for external states.

An external dual point $x_i$ corresponds here to the line $(Z_{i-1}Z_i)$, and a loop point $x_{A_j}$ to the line $(Z_{A_j}Z_{B_j})$, which we abbreviate $(A_jB_j)$. The infinity twistor $I$ is the bitwistor associated with the point at infinity of dual space; it selects the $\lambda$ components of $Z_i=(\lambda_i,\mu_i)$, so that $\langle I\,ij\rangle$ is the ordinary spinor bracket. Writing $\langle ij\rangle\equiv\langle I\,ij\rangle$ and $\langle AB\rangle\equiv\langle I\,A\,B\rangle$ for the two-brackets involving $I$, the dictionary between dual distances and twistor brackets reads
\begin{equation}
x_{ij}^2 \;=\; \frac{\langle i{-}1\,i\,j{-}1\,j\rangle}{\langle i{-}1\,i\rangle\,\langle j{-}1\,j\rangle}\,,\qquad
(x_A-x_i)^2 \;=\; \frac{\langle AB\,i{-}1\,i\rangle}{\langle AB\rangle\,\langle i{-}1\,i\rangle}\,,
\label{eq:dictionary}
\end{equation}
and more generally $(x-y)^2=\frac{\langle PQRS\rangle}{\langle I\,PQ\rangle\langle I\,RS\rangle}$ for points $x,y$ dual to the lines $(PQ)$ and $(RS)$ respectively. The loop integration measure naturally associated with this representation is
\begin{equation}
\int d^4x_{A_j} \;\leftrightarrow\; \int\frac{d^4Z_{A_j}\,d^4Z_{B_j}}{\langle A_jB_j\rangle^4\,\mathrm{vol}\,[GL(2)]}\,.
\label{eq:measure}
\end{equation}
We absorb all factors of $\langle A_jB_j\rangle^{-4}$ into the definition of the integrand, where, as a manifestation of dual conformal invariance, they simplify against analogous terms that arise from the application of eq.~\eqref{eq:dictionary}. In these variables the tree-normalized MHV integrand takes the form
\begin{equation}
\mathcal{I}_{n,0}^{(L)} \;=\; \frac{\mathcal{N}_n^{(L)}\big(Z_1,\dots,Z_n;\,Z_{A_1},Z_{B_1},\dots,Z_{A_L},Z_{B_L}\big)}{\Big(\prod_{i=1}^{n}\prod_{j=1}^{L}\langle i\,i{+}1\,A_jB_j\rangle\Big)\Big(\prod_{1\le r<l\le L}\langle A_rB_rA_lB_l\rangle\Big)}\,.
\label{eq:integrandtw}
\end{equation}
In these variables dual conformal invariance is the requirement that the integrand, including the measure, be invariant under a rescaling of each twistor separately (projective invariance). The numerator must therefore be uniform of degree $2L$ in each external twistor and $(n+L-1)-4$ in each loop twistor, the four missing powers of the latter being supplied by the measure. Planarity further constrains it, since the denominator must simplify against some of its factors so as to leave a sum of planar graphs. For example, the box integrand of eq.~\eqref{eq:box} reads in these variables
\begin{equation}
\mathcal{I}_{4,0}^{(1)} \;=\; -\frac{\langle1234\rangle^2}{\langle AB12\rangle\langle AB23\rangle\langle AB34\rangle\langle AB41\rangle}\,.
\label{eq:boxtw}
\end{equation}

Momentum twistors also simplify the $R$-invariants of section~\ref{sec:symmetries}. The external data become momentum supertwistors $(Z_i|{\chi_i})$, whose Grassmann components ${\chi_i^A}=\lambda_i^\alpha\theta_{i\alpha}^A$ are built from the dual coordinates of eq.~\eqref{eq:dualtheta} and take the place of the on-shell variables of section~\ref{sec:symmetries} (see refs.~\cite{Drummond:2008vq,Mason:2009qx} for details). In these variables the $R$-invariants take the compact form~\cite{Mason:2009qx}
\begin{equation}
[abcde] \;=\; \frac{\delta^{0|4}\big({\chi_a}\langle bcde\rangle+{\chi_b}\langle cdea\rangle+{\chi_c}\langle deab\rangle+{\chi_d}\langle eabc\rangle+{\chi_e}\langle abcd\rangle\big)}{\langle abcd\rangle\,\langle bcde\rangle\,\langle cdea\rangle\,\langle deab\rangle\,\langle eabc\rangle}\,.
\label{eq:Rtw}
\end{equation}
Expanding in a chosen basis $R_a^k$ of invariants of Grassmann degree $4k$, the counterpart of eq.~\eqref{eq:nmhvdual}, and the generalization of eq.~\eqref{eq:integrandtw} to any sector, reads
\begin{equation}
\begin{aligned}
\mathcal{I}_{n,k}^{(L)} \;=\; \sum_a\, &R_a^k(Z_1,\dots,Z_n;\,{\chi_1},\dots,{\chi_n})\\
&\times\frac{\mathcal{N}_{n\,a}^{(L)}\big(Z_1,\dots,Z_n;\,Z_{A_1},Z_{B_1},\dots,Z_{A_L},Z_{B_L}\big)}{\Big(\prod_{i=1}^{n}\prod_{j=1}^{L}\langle i\,i{+}1\,A_jB_j\rangle\Big)\Big(\prod_{1\le r<l\le L}\langle A_rB_rA_lB_l\rangle\Big)}\,,
\end{aligned}
\label{eq:integrandtwk}
\end{equation}
where each $\mathcal{N}_{n\,a}^{(L)}$ is subject to the same weight and planarity requirements as in the MHV case. In every sector the integrand is thus written as numerators over the product of the distinct propagators; the form of the numerators is specified in section~\ref{sec:ansatz}.

\subsection{The internal collinear condition}
\label{app:twistorcond}

The dictionary~\eqref{eq:dictionary} maps the two propagators that vanish in the limit to the adjacent brackets
\begin{equation}
\ell^2=\frac{\langle A_1B_1\,a{-}1\,a\rangle}{\langle A_1B_1\rangle\langle a{-}1\,a\rangle}\,,\qquad
(\ell-p_a)^2=\frac{\langle A_1B_1\,a\,a{+}1\rangle}{\langle A_1B_1\rangle\langle a\,a{+}1\rangle}\,.
\label{eq:twprops}
\end{equation}
On the MHV branch of the double cut the line $(A_1B_1)$ lies in the plane $(a{-}1\,a\,a{+}1)$; on the $\MHVb$ branch it passes through $Z_a$. The collinear point is their intersection, which for $a=1$, the case of eq.~\eqref{eq:intro-condition}, we parameterize as
\begin{equation}
Z_{A_1}=Z_1\,,\qquad Z_{B_1}=\beta Z_n+\gamma Z_2\,.
\label{eq:locus}
\end{equation}
On this locus the internal collinear condition~\eqref{eq:intro-condition} reads
\begin{equation}
\begin{aligned}
\,&\mathcal{I}_{n,k}^{(L)}(Z_1,\dots,Z_n;\,Z_{A_1},Z_{B_1},\dots,Z_{A_L},Z_{B_L};\,{\chi_1},\dots,{\chi_n})\rightarrow\\
&-\frac{1}{\beta\gamma\langle A_1B_1n1\rangle\langle A_1B_112\rangle}\,\mathcal{I}_{n,k}^{(L-1)}(Z_1,\dots,Z_n;\,Z_{A_2},Z_{B_2},\dots,Z_{A_L},Z_{B_L};\,{\chi_1},\dots,{\chi_n})\,.
\end{aligned}
\label{eq:twcond}
\end{equation}
In the MHV sector it is the condition of refs.~\cite{Bourjaily:2011hi,Golden:2012hi}.
On the locus~\eqref{eq:locus} $\langle A_1B_1\rangle=\gamma\langle1\,2\rangle-\beta\langle n\,1\rangle$, and the momentum fraction, defined with the neighboring leg $p_2$ as reference, is
\begin{equation}
z=\frac{p_2\cdot \ell}{p_2\cdot p_a}=\frac{\gamma\,\langle1\,2\rangle}{\langle A_1B_1\rangle}\,,\qquad
1-z=\frac{p_2\cdot(p_a-\ell)}{p_2\cdot p_a}=1-\frac{\gamma\,\langle1\,2\rangle}{\langle A_1B_1\rangle}=-\frac{\beta\,\langle n\,1\rangle}{\langle A_1B_1\rangle}\,.
\label{eq:zfraction}
\end{equation}
Inserting these expressions and the dictionary~\eqref{eq:dictionary} for the two vanishing dual distances in eq.~\eqref{eq:intro-condition} gives eq.~\eqref{eq:twcond} with an additional factor $\langle A_1B_1\rangle^4$ on the right-hand side. This factor simplifies against the factor $\langle A_1B_1\rangle^{-4}$ that arises from the change of measure~\eqref{eq:measure} inside the integration sign.

\subsection{The conditions on external kinematics}
\label{app:twistorext}
\begin{itemize}
\item \emph{Helicity-preserving collinear condition.} The supertwistor $\mathcal{Z}_n=(Z_n|{\chi_n})$ approaches $\mathcal{Z}_{n-1}$,
\begin{equation}
\mathcal{Z}_n=\mathcal{Z}_{n-1}+\delta_c\,\mathcal{Z}_X\,,
\label{eq:hp-collinear}
\end{equation}
with generic $\mathcal{Z}_X$, so that $\langle n{-}1\,n\rangle$ vanishes while $[n{-}1\,n]$ stays finite, and eq.~\eqref{eq:hp-coll-dual} reads
\begin{equation}
\mathcal{I}_{n,k}^{(L)}\big(Z_1,\dots,Z_n\big)=\mathcal{I}_{n-1,k}^{(L)}\big(Z_1,\dots,Z_{n-2},Z_{n-1}\big)+\mathcal{O}(\delta_c)\,.
\label{eq:hp-coll-cond}
\end{equation}
\item \emph{Helicity-changing collinear condition.} Only $[n{-}1\,n]$ vanishes. The pole is a coplanarity,
\begin{equation}
Z_1=Z_n+a\,Z_{n-1}+b\,Z_{n-2}+\delta_c\,Z_X\,,
\label{eq:hc-coll-locus}
\end{equation}
with generic $a$, $b$ and $Z_X$, and fixed ${\chi_1}$. The twistor $\hat Z=a\,Z_{n-1}+b\,Z_{n-2}$ replaces the pair $n{-}1$, $n$, and eq.~\eqref{eq:hc-coll-dual} reads
\begin{equation}
\mathcal{I}_{n,1}^{(L)}\big(Z_1,\dots,Z_n\big)=\mathcal{R}_{n,1}^{\text{tree}}\big(Z_1,\dots,Z_n\big)\Big|_{\delta_c^{-1}}\,\mathcal{I}_{n-1,0}^{(L)}\big(Z_1,\dots,Z_{n-2},\hat Z\big)+\mathcal{O}(\delta_c^0)\,,
\label{eq:hc-coll-cond}
\end{equation}
with the ratio function written through the $R$-invariants~\eqref{eq:Rtw}.
\item \emph{Helicity-preserving soft condition.} The twistor $Z_n$ approaches the line $(Z_{n-1}Z_1)$,
\begin{equation}
Z_n=a\,Z_{n-1}+b\,Z_1+\delta_s\,Z_X\,,
\label{eq:soft-locus}
\end{equation}
with ${\chi_n}=a\,{\chi_{n-1}}+b\,{\chi_1}+\delta_s\,{\chi_X}$, and eq.~\eqref{eq:hp-soft-dual} reads
\begin{equation}
\mathcal{I}_{n,k}^{(L)}\big(Z_1,\dots,Z_n\big)=\mathcal{I}_{n-1,k}^{(L)}\big(Z_1,\dots,Z_{n-1}\big)+\mathcal{O}(\delta_s^2)\,.
\label{eq:hp-soft-cond}
\end{equation}
\item \emph{Helicity-changing soft condition.} On the locus~\eqref{eq:soft-locus} with ${\chi_n}$ fixed, the $R$-invariants that contain $Z_{n-1}$, $Z_n$ and $Z_1$ have a double pole, and eq.~\eqref{eq:hc-soft-dual} reads
\begin{equation}
\mathcal{I}_{n,1}^{(L)}\big(Z_1,\dots,Z_n\big)=\mathcal{R}_{n,1}^{\text{tree}}\big(Z_1,\dots,Z_n\big)\Big|_{\delta_s^{-2}}\,\mathcal{I}_{n-1,0}^{(L)}\big(Z_1,\dots,Z_{n-1}\big)+\mathcal{O}(\delta_s^{-1})\,.
\label{eq:hc-soft-cond}
\end{equation}
\end{itemize}

\bibliographystyle{JHEP}
\bibliography{v20}
\end{document}